\documentclass[a4paper,11pt]{article}
\pdfoutput=1 

\usepackage{jheppub} 

\usepackage[T1]{fontenc} 

\usepackage{slashed}
\usepackage{multirow}
\usepackage{enumerate}
\usepackage{enumitem}
\usepackage{float}
\usepackage{ulem}

\usepackage[utf8]{inputenc}
\usepackage{amssymb}
\usepackage{amsmath}
\usepackage{amsfonts}
\usepackage{graphicx}
\usepackage{color}
\usepackage{xspace}
\usepackage{comment}
\usepackage[section]{placeins}
\usepackage{afterpage}
\usepackage{calc}
\usepackage{caption}
\usepackage{subfigure}
\usepackage{verbatim}
\usepackage{multirow,rotating}
\usepackage[dvipsnames]{xcolor}
\usepackage{graphicx}
\usepackage{mathrsfs}

\newcommand{\be}{\begin{equation}}
\newcommand{\ee}{\end{equation}}
\newcommand{\bea}{\begin{eqnarray}}
\newcommand{\eea}{\end{eqnarray}}

\newcommand{\met}{\slashed{E}_T}

\newcommand{\iab}{\rm ab^{-1}}
\newcommand{\ifb}{\rm fb^{-1}}
\newcommand{\mltp}{{\mkern-2mu\times\mkern-2mu}}
\newcommand{\iTeV}{\,{\mathrm{{TeV}^{-1}}}}

\title{\boldmath Probing axion-like particles coupling to gluons at the LHC}

\author[a]{Filmon Andom Ghebretinsae}
\author[b,c]{, Zeren Simon Wang}
\author[a]{and Kechen Wang}

\affiliation[a]{Department of Physics, School of Science, Wuhan University of Technology, 430070 Wuhan, Hubei, China}
\affiliation[b]{Department of Physics, National Tsing Hua University,	Hsinchu 300, Taiwan}
\affiliation[c]{Center for Theory and Computation, National Tsing Hua University,	Hsinchu 300, Taiwan}

\emailAdd{filmonandom@whut.edu.cn}
\emailAdd{wzs@mx.nthu.edu.tw}
\emailAdd{kechen.wang@whut.edu.cn}

\abstract{
Assuming ALPs couple to gluons only, they can be produced via the $p p \to a j$ process and decay into two jets at the LHC. When the coupling parameter, $C_{\tilde{G}} / f_a$, is small, the lifetime of ALPs can be long enough leading to displaced final state jets. In this paper, we consider the signal including both the prompt and long-lived cases of ALPs by employing a specialized Delphes module to handle displaced jets. Relevant background processes are generated and simulated at the detector level, and multivariate analyses based on machine-learning are performed to discriminate signal and background events and achieve the best sensitivities. Based on the data accumulated for this study, we forecast the expected upper limits on $C_{\tilde{G}}/f_a$ for ALP mass $m_a$ in the range 5$-$2300 GeV at 2-, 3- and 5-$\sigma$ significances at the High Luminosity-LHC with 14 TeV center-of-mass energy and 3 $\iab$ integrated luminosity. Vast previously unprobed regions in the parameter space spanned by $C_{\tilde{G}}/f_a$  and $m_a$ are probed and the best upper limits on $C_{\tilde{G}}/f_a$ at 2-$\sigma$ significance is found to be around $1.0 \times 10^{-2} \,\, {\rm TeV^{-1}}$ for $m_a \sim 500$ GeV. The ALP mass is reconstructed from the kinematics of final state jets and we find that it is measurable in this method when $m_a$ is below about 1 TeV at the HL-LHC. The effects of systematic uncertainties and validation of the EFT framework are also checked and discussed.
}

\begin{document} 
\maketitle
\flushbottom

\section{Introduction}
\label{sec:Intro}

Granting the fact that the Standard Model (SM) has demonstrated huge successes in providing experimental predictions, and is generally believed to be a self-consistent effective theory, it still leaves some phenomena unexplained and is deprived of being a complete theory of everything. 
One of such challenges is the strong CP problem in QCD~\cite{Dine:2000cj,Kim:2008hd,Hook:2018dlk}.
The resolution to the strong CP problem came through the works of Peccei and Quinn \cite{PhysRevLett_38_1440, Peccei:1977ur} by incorporating a spontaneously broken global U$(1)_{\rm PQ}$ symmetry in the SM,
 and axions are the pseudo-Nambu-Goldstone bosons of U$(1)_{\rm PQ}$.
Over the decades, several aspects of the original axion model have been investigated and more models have been developed~\cite{Weinberg:1977ma, Wilczek:1977pj, Kim:1979if, Shifman:1979if, Zhitnitsky:1980tq, Dine:1981rt, Hook:2019qoh,Quevillon:2019zrd, DiLuzio:2020wdo}.

In the original model, 
the mass of QCD axion is strictly related to the energy scale $f_{a}$ of the global symmetry breaking. Setting the stringent relation between the axion mass and $f_{a}$ aside, the idea of Axion-like Particles (ALPs) with wide-ranging masses and couplings arises. 
Unlike QCD axions, the couplings and masses of ALPs are considered to be independent parameters. 
Generally, these pseudo-Nambu-Goldstone bosons can arise naturally from the spontaneous breaking of a new U(1) global symmetry at some large energy scale $f_a$ in many beyond standard model theories.
They are gauge-singlets, odd under CP transformations,
and the strength of their couplings to SM particles is proportional to the inverse of the global symmetry breaking scale $f_{a}$.

Because their masses and couplings to SM particles can range over many orders of magnitude, 
ALPs have been searched extensively according to different production processes and decay products at colliders and constraints are set in different regions of parameter space spanned by the ALP mass and various couplings.
Collider studies on ALPs can be found in the reviews~\cite{Bauer:2017ris, Dolan:2017osp, Bauer:2018uxu, Zhang:2021sio, dEnterria:2021ljz, Agrawal:2021dbo} and references therein.
Recent collider experimental searches for ALPs by the ATLAS~\cite{ATLAS:2015rsn, ATLAS:2018coo, ATLAS:2021hbr}, CMS~\cite{CMS:2012qms, CMS:2015twz, CMS:2017dmg, CMS:2019ruu, CMS:2020bni}, and Belle II~\cite{Belle-II:2020jti} collaborations are also reviewed in Ref.~\cite{Tian:2022rsi}.

Current studies mainly focus on probing ALPs' couplings to electroweak gauge bosons (photons, $Z$- and $W$-bosons), SM Higgs bosons, and leptons, while the coupling to gluons needs more investigations, especially for heavy ALPs.
Ref.~\cite{Blinov:2021say} reviews some current constraints on the ALP-gluon coupling from various collider experiments. 
These include 
GlueX constraints through the photoproduction of ALPs \cite{GlueX:2021myx}, and 
reinterpretations from KOTO search for $K_L\rightarrow\pi^0$ + invisible \cite{KOTO:2018dsc}, NA62 search for $K^{+} \rightarrow \pi^{+}$ + invisible \cite{NA62:2021zjw}, NA62 search for $K^{+} \rightarrow \pi^{+} a (\gamma\gamma)$ \cite{NA62:2014ybm}, proton beam-dump searches at the $\nu$CAL \cite{Blumlein:1990ay, Blumlein:1991xh} and CHARM \cite{CHARM:1985anb}, 
and  constraints on rare pion decay $\pi^{+}\rightarrow ae\nu$ at the  PIENU \cite{PIENU:2017wbj} and PIBETA \cite{Pocanic:2003pf}.
All of the above studies set limits on ALP-gluon coupling for ALP masses below about 1 GeV.
Moreover, Ref.~\cite{Aloni:2018vki} reviews existing constraints on the ALP-gluon coupling in the mass range $m_{\pi} < m_a < 3$ GeV, and more current limits on the ALP-gluon coupling can be found in review~\cite{Bauer:2021mvw} and references therein.

Concerning collider searches for constraining the ALP-gluon-gluon coupling at the LHC,
the ATLAS collaboration has performed one study through the decay of the Higgs boson into the $Z$-boson and an ALP ($h \to Z\, a$), in which the ALP decays into a gluon pair ($a \to g g$)~\cite{ATLAS:2020pcy}.
The ATLAS collaboration has also probed ALPs' coupling to gluons through the mono-jet final-state topology \cite{ATLAS:2021kxv}, i.e. an energetic jet plus large missing transverse momentum,  using proton-proton collision data with a center-of-mass energy $\sqrt{s} =$ 13 TeV and an integrated luminosity $\mathcal{L}_{\rm int} = $  139 $\ifb$.
The study has explored ALP masses up to $m_a$ = 1 GeV, ALP-gluon couplings up to $C_{\tilde{G}}$ = 1 and effective scales $f_a$ in the range 1-10 TeV. 
The result has excluded the values of $C_{\tilde{G}}/f_a$ above $8\times10^{-3} \, {\rm TeV^{-1}}$ at 95\% confidence level, and the exclusion does not significantly depend on the ALP masses up to about 1 GeV.

Another search constrains the ALP-gluon-gluon coupling through the signal process $ g g \to a \to ZZ, Zh$ at the LHC by the CMS collaboration~\cite{CMS:2021xor}.
Furthermore, the CMS collaboration has carried out a search for a vector resonance produced in proton-proton collision and decaying into a quark-antiquark pair, accompanied by a high $p_T$ jet from initial state radiation (ISR)~\cite{CMS:2017dcz}. 
The minimum $p_T$ of the vector resonance is high enough, so that its decay products are merged into a single, large-radius jet. 
They analyze the data with $\sqrt{s} =$ 13 TeV and $\mathcal{L}_{\rm int} = $ 35.9 $\ifb$, and set upper limits on the production cross section  for the vector boson mass in the range of 50-300 GeV.
The results are reinterpreted to constrain the ALP-gluon coupling in Ref. \cite{Mariotti:2017vtv}, by recasting the limits on the production cross section of a $q\bar{q}$-initiated resonance $\sigma_{q\bar{q}}^{\rm CMS}$ given in Ref.~\cite{CMS:2017dcz} to that of a $g g$-initiated resonance $\sigma_{gg}$ through the equation $\sigma_{gg} = \sigma_{q\bar{q}}^{\rm CMS} \cdot {\epsilon_{H_T}^{q\bar{q}}} / {\epsilon_{H_T}^{gg}}$. 
Here, $\epsilon_{H_T}^{q\bar{q}}$ and $\epsilon_{H_T}^{gg}$ are cut efficiencies when applying hadronic activity $H_T > $ 650 GeV.
Converting into our notation convention of ALP-gluon coupling $C_{\tilde{G}}/f_a$ in following sections, 
Ref. \cite{Mariotti:2017vtv} shows that $C_{\tilde{G}}/f_a \gtrsim 3\times10^{-2}\,\, {\rm TeV^{-1}}$ can be excluded for ALP masses $m_a$ in the range 50-125 GeV.

Additionally, a plethora of phenomenological researches~\cite{Kleban:2005rj, Mimasu:2014nea, CidVidal:2018blh, Curtin:2018mvb, Ebadi:2019gij, Aloni:2019ruo, Goncalves:2020bqi, Gershtein:2020mwi, Haghighat:2020nuh, Knapen:2021elo} have also been amassed on the sensitivities of the ALP-gluon-gluon coupling, albeit mostly focusing on the ALP mass range below a few GeVs.
Recently, we have presented a strategy to detect the long-lived ALPs and explore the discovery potential of various FAr Detectors at the Electron Positron Collider (FADEPC)~\cite{Tian:2022rsi}, where
we estimate their sensitivities on the model parameters in terms of the effective ALP-photon-photon coupling, the effective ALP-photon-$Z$ coupling, and ALP mass $m_a$ via the signal process $e^-e^+ \rightarrow  \gamma \,\, a,~ a \to \gamma\gamma $  at future $e^{-}e^{+}$ colliders running at center-of-mass energy of $\sqrt{s} = 91.2$ GeV and integrated luminosities of 16, 150, and 750 ab$^{-1}$.

In this work, we probe the ALP's coupling to gluons at the High-Luminosity Large Hadron Collider (HL-LHC) via the signal process $p p \to a j, a \to j j$.
Since this process can be independent of ALP's couplings to other SM particles, it is a unique signal channel if ALP couples to gluons only.
The signal including both the prompt and long-lived cases of ALPs are considered and multivariate analyses based on machine-learning are performed to reject background events and achieve the best sensitivities.
The expected upper limits on $C_{\tilde{G}}/f_a$ for ALP mass in the range 5$-$2300 GeV are derived at the HL-LHC with 14 TeV center-of-mass energy and 3 $\iab$ integrated luminosity.

This article is structured as follows: 
Sec.~\ref{sec:SigProd} describes the theoretical foundation and production cross section for the signal process. 
In Sec.~\ref{sec:Sear_Stra}, we present the search strategy including the relevant SM background processes, the methodology employed for the event simulation, and the data analyses thoroughly. 
Sec.~\ref{sec:Results} shows the sensitivities on the model parameters achieved in this study.
The method to reconstruct the ALP mass, the effects of systematic uncertainties and the validation of the effective Lagrangian are also discussed in this section.
We summarize and conclude in Sec.~\ref{sec:Conc}. 
Some details of our study which complement the main body of the paper are listed in the appendices.

\section{The Signal Production}
\label{sec:SigProd}

Assuming ALPs couple to gluons only,
the effective linear ALP Lagrangian is given by~\cite{Brivio:2017ije}

\begin{equation}
\begin{aligned}
{\mathscr L}_{\rm eff}
\supset \,\frac{1}{2}\,(\partial_{\mu}a)\,(\partial^{\mu}a) - \,\frac{1}{2}\,m^2_{a}\,a^2 - \, \frac{C_{\tilde{G}}}{f_a} \, a\,G^A_{\mu\nu}\,\tilde{G}^{\mu\nu,A}\, ,
\label{eqn:Eff_Lag} 
\end{aligned}
\end{equation}
where $G^A_{\mu\nu}$ ($A$ = 1,...,8) represents the gluon field strength tensor corresponding to $\rm SU_c (3)$ group and $\tilde{G}^{\mu\nu,A} \equiv \frac{1}{2}\varepsilon^{\mu\nu\alpha\beta}G^A_{\alpha\beta}$ is the dual field strength tensor.
$m_a$ and $f_a$ represent the ALP mass and the characteristic energy scale of the global symmetry breaking, correspondingly, and are assumed to be independent parameters throughout this article.

\begin{figure}[h]
\centering
\includegraphics[width=10cm]{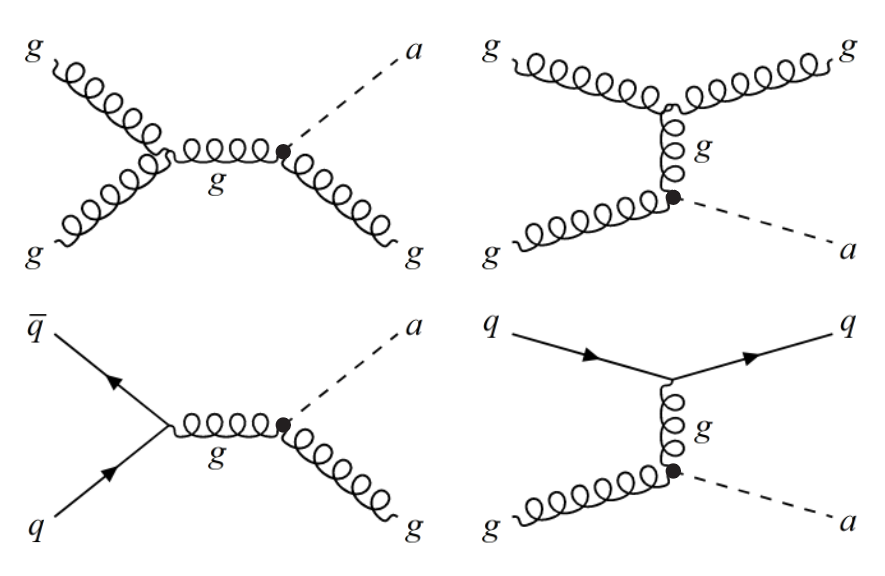}
\caption{
Representative production processes of ALPs in association with a jet at $pp$ colliders ($p \, p \,\rightarrow a \, j$) when only ALP's coupling to gluons is considered.
}
\label{fig:Feyn_Diag}
\end{figure}

In this study, we consider the production of ALPs in association with a jet at the HL-LHC with a center-of-mass energy $\sqrt{s}$ = 14 TeV and integrated luminosity $\mathcal{L}_{\rm int}$ = 3 $\iab$. The corresponding signal process is, therefore, $p \, p \,\rightarrow a \, j$ and
representative processes are represented in Fig.~\ref{fig:Feyn_Diag}.

\begin{figure}[h]
\centering
\includegraphics[width=10cm, height=6cm]{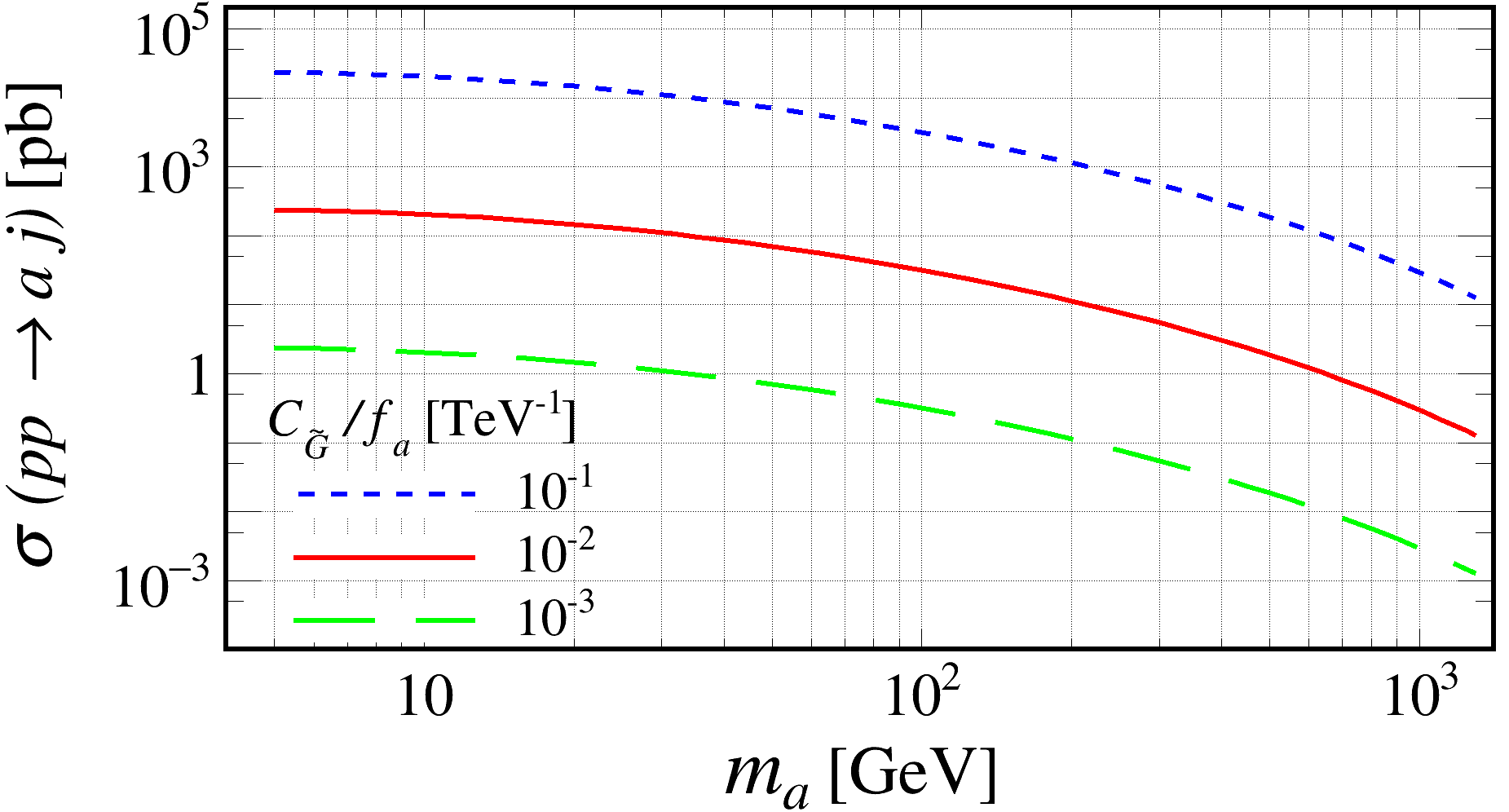}  
\caption{
Signal production cross section $\sigma(p \, p \,\rightarrow a \, j)$ as a function of ALP mass $m_a$ and the parameter $C_{\tilde{G}}/f_a$. 
}
\label{fig:Cross_vs_ma}
\end{figure}

We implement the ALP model file with the linear Lagrangian~\cite{Brivio:2017ije} in the Universal FeynRules Output (UFO) format~\cite{Degrande:2011ua} into the \textsc{MadGraph5} program~\cite{Alwall:2014hca} to simulate the proton-proton collisions and generate the $p \, p \,\rightarrow a \, j$ events. Details of data simulation are shown in Sec.~\ref{subsec:Data_Sim}.
To maintain consistency throughout our study, the production cross sections calculated by \textsc{MadGraph5} are used to estimate the number of signal events.
Fig.~\ref{fig:Cross_vs_ma} shows the dependence of the production cross section of the signal process, $\sigma(p \, p \,\rightarrow a \, j)$, on the model parameters $m_a$ and $C_{\tilde{G}}/f_a$. 
As we can see from the figure, the production cross section declines with the ALP mass $m_a$ for fixed $C_{\tilde{G}}/f_{a}$, 
while it is proportional to the square of $C_{\tilde{G}}/f_{a}$.
In this paper, we have focused on the interactions of the ALPs with gluons. As such, we assume that the ALP decays completely into a pair of gluons with a decay width of 
\begin{equation}
\begin{aligned}
\Gamma(a \, \rightarrow \, gg) = \frac{2}{\pi} m_a^3 \left| \frac{C_{\tilde{G}}}{f_a} \right|^2
\label{eqn:Decay_Width}
\end{aligned} 
\end{equation}
at leading order~\cite{Bauer:2017ris, Aloni:2018vki} and a branching ratio Br($a \rightarrow g g$) = 1 \footnote{Assuming ALP couples to gluons only, for the considered ALP mass range of $m_a > 5$ GeV in this article, ALP decays dominantly into a gluon pair~\cite{Bauer:2017ris}, while it decays to mesons when $m_\pi \lesssim m_a \lesssim 3$ GeV~\cite{Aloni:2018vki}.}.

\vspace{0 cm}
\section{Search Strategy}
\label{sec:Sear_Stra}

\subsection{The SM Background}
\label{subsec:SM_BKG}

The dominant background processes that are relevant to our signal are as follows.
\begin{enumerate}[label*=(\roman*)]
\item multi-jet production;
\item A pair of taus in association with a jet, $\tau^+\tau^-j$; 
\item Production of massive gauge bosons in association with a jet, $W^\pm j$ and $Z j$.
\item Production of two massive gauge bosons, $W^+ W^-$ and $ZZ$.
\item Top quark pair production, $t\bar{t}$.
\end{enumerate}

At proton-proton colliders, multi-jet background has very large cross section and , hence, is more significant than other SM background processes. Therefore, it requires large event samples during simulation. 
In this study, because the signal final state has at least three jets and the computational resource is limited, we have generated totally $1.244 \times 10^8$ events of $p p \to j j$ and $ j j j$ processes and combine them as multi-jet samples.
The remaining background processes have relatively smaller cross sections, and 
$10^6\,\, \tau^+\tau^-j$, $3 \times 10^5\,\, W^\pm j$, $3 \times 10^5\,\, Z j$, $10^5 \,\, W^+ W^- $, $10^5\,\, ZZ$ and $10^5\,\, t\bar{t}$ events have been generated.

\subsection{Event Simulation}
\label{subsec:Data_Sim}

First, events are generated using \textsc{madgraph5{\textunderscore}}a\textsc{mc}@\textsc{NLO} version 2.6.7~\cite{Alwall:2014hca} using the \textsc{nnpdf23} \cite{Ball:2012cx} as the parton distribution function (PDF) of the proton at 14 TeV center-of-mass energy.
For the signal, a scan of the ALP mass, $m_{a}$, is done in the following fashion: 15 GeV increments in the mass range 5$-$50 GeV, 25 GeV increments in the range 50$-$100 GeV, 50 GeV increments in the range 100$-$500 GeV and 200 GeV increments in the range 500$-$2300 GeV. For each ALP mass, the second model parameter, $C_{\tilde{G}}/f_a$, has been scanned between $1.0 \times 10^{-2} - 5.5 \times 10^{-2} \,\,{\rm TeV^{-1}}$ with $5.0 \times 10^{-3}\,{\rm TeV^{-1}}$ increments for masses up to 1300 GeV and between $1.0 \times 10^{-2} - 2.5 \times 10^{-1}\,\,{\rm TeV^{-1}}$ with coarse steps for all the masses above 1300 GeV. For each iteration of ALP mass $m_a$ and coupling $C_{\tilde{G}}/f_a$, $10^6$ events have been generated to ensure that statistical uncertainties are minimized as much as possible within the bounds of computational resources.

At the parton level, all signal and background events are required to satisfy the following relatively lenient cuts: 
(a) The minimum transverse momentum of jets and b-jets is set to 5 GeV, i.e. $p_{T}(j/b)>$ 5 GeV, and for photons and charged leptons it is set to 1 GeV, i.e. $p_{T}(\gamma/l^{\pm}) > 1$ GeV;
(b) The maximum pseudorapidity of all particles is set to 8, i.e. $|\eta(j/b/\gamma/l^{\pm})| < 8$; 
(c) the minimum angular separation between every permutation of jets, b-jets, photons and charged leptons is set to 0.01, i.e. $\Delta R(j/b/\gamma/l^{\pm},j/b/\gamma/l^{\pm}) > 0.01$.

All events are then passed through \textsc{pythia} version 8.244~\cite{Sjostrand:2014zea} for parton showering, hadronization and decay of unstable particles. 
For detector effects, we use the fast detector simulation tool \textsc{delphes}~\cite{deFavereau:2013fsa} with an ATLAS detector configuration. 
The FastJet version 3.1.3~\cite{Cacciari:2011ma} package is used to perform jet reconstruction employing the anti-$k_{t}$ algorithm \cite{Cacciari:2008gp} with a jet cone size of 0.3  and a minimum transverse momentum of 5 GeV.
Identifying displaced vertices in \textsc{delphes} has been very tricky and an official workable module has not yet been released. 
However, Ref.~\cite{Nemevsek:2018bbt}  introduces a customized version, \textsc{delphes} version 3.3.2\_JD\_TVS~\cite{Delphes:wDisp}, including additional modules such as TrackVertexSmearing and JetDisplacement modules that allow the definition of a displaced jet. More specifically, the transverse displacement of a jet $V_{T} (j) = \sqrt{d_{x}^{2} (j) + d_{y}^{2} (j)}$, where $d_x$ and $d_y$ are displacements of the vertex along the respective axes, is defined by taking into consideration the displacement of the origin of all tracks with transverse momentum larger than a certain threshold and selecting the minimum of this displacements. For our analysis, we employed the use of these displaced jet modules and set $\Delta R ({\rm track, j})<$ 0.3 and $p_T ({\rm track}) >$ 20 GeV.

\subsection{Data Analyses}
\label{subsec:Analyses}

At the detector level, the following pre-selection cuts have been employed before a multivariate analysis is done.
\begin{enumerate}[label*=(\arabic*)]
\item We require that each event has at least three jets, i.e. $N (j) \geq$ 3 and events that contain electrons, muons and photons are vetoed, i.e. $N(e) = N (\mu) = N (\gamma) = $ 0.
\item Events containing b-tagged and tau-tagged jets are vetoed, i.e. $N(b) = N(\tau) = $ 0.
\item The minimum transverse momentum of the first three hardest jets is set to 30 GeV, i.e. $p_T (j_1)$, $p_T (j_2)$, $p_T (j_3) >$ 30 GeV. \label{last-item}
\item The maximum missing transverse energy is set to 30 GeV, i.e. $\met <$ 30 GeV. 
\end{enumerate}

The expected number of events $N_{\rm exp} = \sigma_{\rm prod} \times \mathcal{L}_{\rm int} \times \epsilon_{\rm pre}$ after applying pre-selection cuts (1)-(4) sequentially are displayed in Table~\ref{tab:Crsc}. Here, $\sigma_{\rm prod}$ is the production cross section of the signal/background process, $\mathcal{L}_{\rm int}$ is the integrated luminosity, which we have taken to be 3 $\iab$ for the LHC, and $\epsilon_{\rm pre}$ is the pre-selection cut efficiency. 
One sees that the multi-jet background is still very large after all pre-selection cuts and is a factor of $\sim 10^3$ larger than the other background processes. Therefore, the background is dominated by the multi-jet process.
After pre-selection, the jets are sorted according to their transverse displacement $V_T$, therefore $j_1$ beyond this point refers to the most displaced jet in the event, $j_2$ the second most displaced jet and so forth.
\begin{table}[h]
\centering
\begin{tabular}{cccccc}
\hline 
\hline
& initial  & cut(1) &  cut(2) & cut(3) & cut(4) \\ 	
\hline 
Signal & $5.58 \mltp 10^{6}$ & $5.55 \mltp 10^{6}$ & $3.72 \mltp 10^{6}$ & $3.37 \mltp 10^{6}$ & $2.45 \mltp 10^{6}$ \\
\hline 
multi-jet & $3.30 \mltp 10^{17}$ & $9.53 \mltp 10^{16}$ & $8.96 \mltp 10^{16}$ & $1.14 \mltp 10^{14}$ & $1.04 \mltp 10^{14}$ \\ 
$W^\pm j$ & $4.35 \mltp 10^{11} $ & $4.18 \mltp 10^{11}$ & $3.66 \mltp 10^{11}$ & $3.27 \mltp 10^{10}$ & $3.02 \mltp 10^{10}$ \\ 
$Zj$ & $1.39 \mltp 10^{11}$ & $1.33 \mltp 10^{11}$ & $9.93 \mltp 10^{10}$ & $1.18 \mltp 10^{10}$ & $1.08 \mltp 10^{10}$ \\
$\tau^+\tau^- j$ & $3.02 \mltp 10^{10}$ & $1.55 \mltp 10^{10}$ & $1.16 \mltp 10^{10}$ & $1.06 \mltp 10^{8}$ & $6.59 \mltp 10^{7}$ \\
$t\bar{t}$ & $1.84 \mltp 10^{9}$ & $1.80 \mltp 10^{9}$ & $1.56 \mltp 10^{8}$ & $1.45 \mltp 10^{8}$ & $1.09 \mltp 10^{8}$ \\
$W^+W^-$ & $2.16 \mltp 10^{8}$ & $2.13 \mltp 10^{8}$ & $1.72 \mltp 10^{8}$ & $7.29 \mltp 10^{7}$ & $6.50 \mltp 10^{7}$ \\
$ZZ$ & $3.16 \mltp 10^{7}$ & $3.10 \mltp 10^{7}$ & $1.79 \mltp 10^{7}$ & $9.31 \mltp 10^{6}$ & $8.19 \mltp 10^{6}$ \\	
\hline 
\hline
\end{tabular} 
\caption{The expected number of events for the signal (with benchmark mass $m_a$ = 500 GeV and coupling $C_{\tilde{G}}/f_a = 10^{-2} \,\, {\rm TeV^{-1}}$) and relevant SM background processes after applying pre-selection cuts (1)-(4) sequentially at the HL-LHC with center-of-mass energy 14 TeV and integrated luminosity 3 $\iab$.
}
\label{tab:Crsc}
\end{table}

Next, the Toolkit for Multivariate Analysis (TMVA) package~\cite{Hocker:2007ht} is used to perform the Multivariate analyses (MVA) and better discriminate signal and background events. 
The events which pass pre-selection cuts are analyzed through the Boosted Decision Trees (BDT) algorithm in the TMVA package where the discrimination is done using the following thirty observables.

\begin{enumerate}[label=(\alph*)]
\item Missing transverse energy: $\met$.
\item The sum of the transverse momenta of all final state jets: $H_T = \sum_{i}^{N(j)} |\vec{p}_T|$.
\item Transverse momentum, pseudorapidity and azimuth angle of the first three jets: $p_T (j_1)$, $p_T (j_2)$, $p_T (j_3)$; $\eta (j_1)$, $\eta (j_2)$, $\eta (j_3)$; $\phi (j_1)$, $\phi (j_2)$, $\phi (j_3)$.
\item Transverse and longitudinal displacements of the first three jets: $V_T (j_1)$, $V_T (j_2)$, $V_T (j_3)$; $V_z (j_1)$, $V_z (j_2)$, $V_z (j_3)$.
\item Pseudorapidity difference, azimuth angle difference and angular separation between the first and second jets: $\Delta\eta (j_1, j_2)$, $\Delta\phi (j_1, j_2)$, $\Delta R (j_1, j_2)$.
\item Invariant mass, transverse momenta, pseudorapidity and azimuth angle of the system of the first and second jets: $M(j_1 + j_2)$, $p_T (j_1 + j_2)$, $\eta (j_1 + j_2)$, $\phi (j_1 + j_2)$.
\item Transverse momentum, pseudorapidity and invariant mass of the system of the first three jets: $p_T (j_1 + j_2 + j_3)$ - $\eta (j_1 + j_2 + j_3)$, $M(j_1 + j_2 + j_3)$.
\item Pseudorapidity difference, azimuth angle difference and angular separation between the system of $(j_1+j_2)$ and $j_3$: $\Delta\eta(j_1 + j_2, j_3)$, $\Delta\phi(j_1 + j_2, j_3)$, $\Delta R (j_1 + j_2, j_3)$.
\end{enumerate}

\begin{figure}[h]
\centering
\includegraphics[width=10cm, height=6cm]{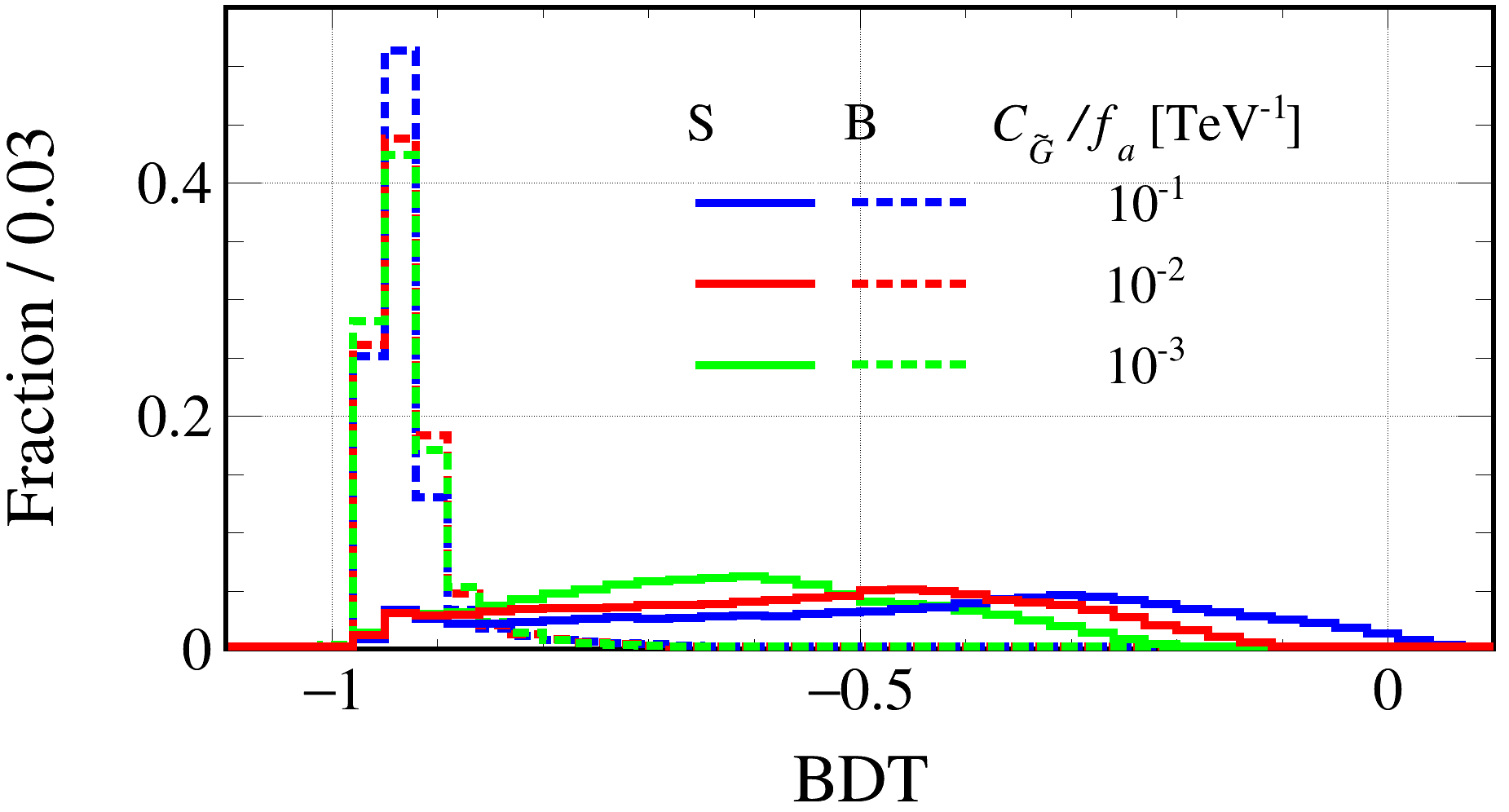}
\caption{
BDT response distributions for total SM background and the signal with ALP mass $m_a$ = 500 GeV and three different ALP-gluon couplings, $C_{\tilde{G}}/f_a$ = $10^{-1}$, $10^{-2}$ and $10^{-3}\,\, {\rm TeV}^{-1}$ at the HL-LHC with center-of-mass energy of 14 TeV.
}
\label{fig:BDT_TBG}
\end{figure}

Fig.~\ref{fig:Input_Disc_Var} in Appendix~\ref{append:BDT_Dis} represents distributions of representative variables used as input to the multivariate analysis for the benchmark mass $m_{a}$ = 500 GeV and parameter $C_{\tilde{G}}/f_a = 10^{-2}\,\, {\rm TeV}^{-1}$ at the HL-LHC.
In Fig.~\ref{fig:BDT_TBG}, we present BDT response distributions for total SM background and the signal with ALP mass $m_a$ = 500 GeV and three different ALP-gluon couplings, $C_{\tilde{G}}/f_a$ = $10^{-1}$, $10^{-2}$ and $10^{-3}\,\, {\rm TeV}^{-1}$ at the HL-LHC with center-of-mass energy of 14 TeV. It can be seen that there is no significant difference in the spread of the background BDT score with the change in $C_{\tilde{G}}/f_a$ but the signal distribution spreads away from the background as $C_{\tilde{G}}/f_a$ increases, or in other words, the signal and background discrimination becomes more obvious with the increase in $C_{\tilde{G}}/f_a$ for a fixed $m_a$.

Given the specified method for the multivariate analysis, 
the TMVA package ranks the input observables according to their importance when discriminating between the signal and background.
As the ALP mass varies, the signal kinematics changes, and therefore, the rank also changes.
In Fig.~\ref{fig:Input_Disc_Var_masses} of Appendix~\ref{append:Obsrank}, we show 
distributions of observables for the signal with benchmark coupling $C_{\tilde{G}}/f_a = 10^{-2} \,\, {\rm TeV^{-1}}$ and representative ALP masses, and for the total SM background.
For $m_a \gtrsim 900$ GeV, $H_T$, $p_T(j_1)$ and $p_T(j_2)$ are the highest ranked observables.
This is because as ALP masses get higher, the jets from the decay of the ALPs are harder, and therefore, these observables related to the jet momentum become more distinct between the signal and background, as demonstrated through the $m_a =$ 1500 GeV case in Fig.~\ref{fig:Input_Disc_Var_masses}. 
For $m_a \lesssim $ 100 GeV, since the jets from ALP decay are soft, the distributions of jet momentum-related kinematic variables are similar between the signal and background, as demonstrated through the $m_a =$ 50 GeV case in Fig.~\ref{fig:Input_Disc_Var_masses}.
Therefore, these observables are not as discriminating as in the case of heavier ALPs, while angular observables of final state jets play a more important role.
$\phi(j_{1})$ and $\phi(j_{2})$ are the highest ranked observables in this mass range.
For 100 GeV $\lesssim m_a \lesssim $ 900 GeV, both momentum-related and angular variables can have similar discriminating powers.
Therefore, $H_T$, $p_T(j_{i})$ and $\phi(j_{i})$ are alternatively ranked highest in this mass range.

\section{Results}
\label{sec:Results}

\subsection{Sensitivities on Model Parameters}
\label{subsec:Limits}

As has been mentioned, multivariate analysis has been used to  further discriminate the signal and background processes, and this analysis allots a BDT score for each of the processes. In the next part of our analysis, we use the following formula to calculate the statistical significance,
\begin{equation}
\sigma_{\rm stat} = \sqrt{2 \left[\left(N_s+N_b\right) {\rm ln}\left(1+\frac{N_s}{N_b}\right) - N_s \right] },
\label{eqn:Stat_Sgf}
\end{equation}
where $N_s$ and $N_b$ are the expected numbers of signal and total background events, respectively, after all the selection cuts.
The method we employ for finding the discovery limits in the parameter space spanned by the parameter $C_{\tilde{G}}/f_a$ and the ALP mass $m_{a}$ is as follows.

Fig.~\ref{fig:BDT_Dis} in Appendix~\ref{append:BDT_Dis} depicts BDT response distributions for the signal (black, filled) with fixed parameter $C_{\tilde{G}}/f_a = 10^{-2} \,\, {\rm TeV^{-1}}$ and dominant SM background processes at the HL-LHC assuming different $m_a$ cases.
As we can see from the distributions, as the ALP mass increases the BDT responses for the signal and background become more distinct. This has the advantage of rejecting the background while retaining the signal events when applying a BDT cut. However, as shown in Fig.~\ref{fig:Cross_vs_ma}, the signal production cross section decreases rapidly as the ALP mass increases, which limits the discovery potential for heavy ALP.

\begin{figure}[h]
\centering
\includegraphics[width=10cm, height=6cm]{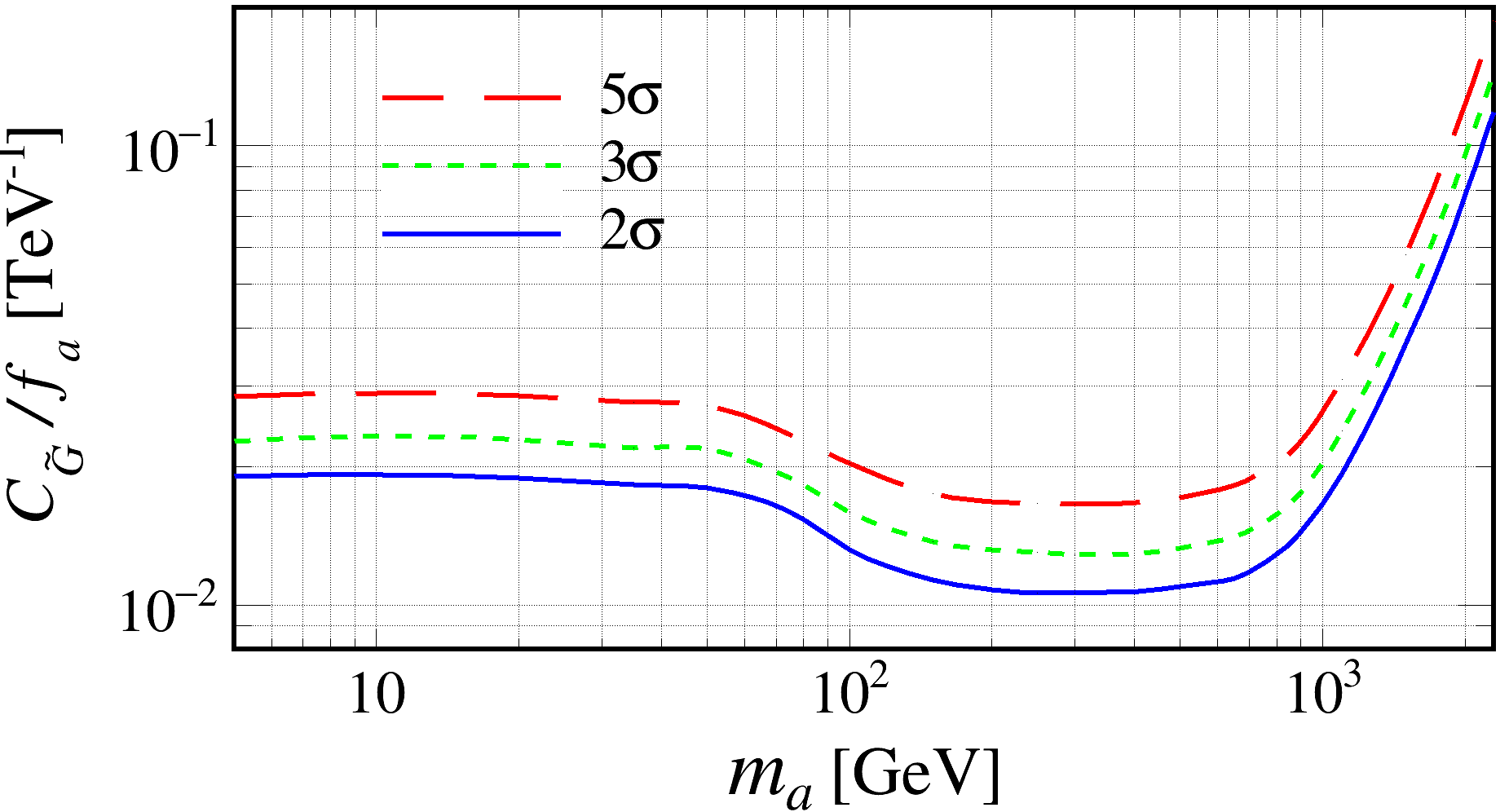}  
\caption{
Upper limits on the model parameter $C_{\tilde{G}}/f_a$ for ALP mass $m_a$ in range 5-2300 GeV at 2-$\sigma$ (solid, blue), 3-$\sigma$ (short dashed, green) and 5-$\sigma$ (long dashed, red) significances at the HL-LHC with a center-of-mass energy of 14 TeV and integrated luminosity 3 $\iab$ . 
}
\label{fig:Upper_Limit}
\end{figure}

For a case with fixed ALP mass $m_a$ and parameter $C_{\tilde{G}}/f_a$,  a BDT cut is then chosen to further reject the background.
Table~\ref{tab:All_Eff} in Appendix~\ref{append:Sel_Eff} shows selection efficiencies for the signal processes with representative ALP masses and fixed $C_{\tilde{G}}/f_a = 10^{-2} \,\, {\rm TeV^{-1}}$ and the background processes at the HL-LHC. 
The final upper limits corresponding to 2-, 3- and 5-$\sigma$ significances at the HL-LHC with center-of-mass energy $\sqrt{s}=14$ TeV and an integrated luminosity of 3 $\iab$ are shown in Fig.~\ref{fig:Upper_Limit}.

\subsection{Reconstruction of Invariant Mass of the ALPs}
\label{subsec:Recon}

As has been mentioned previously, in this study we have assumed that ALPs couple only to gluons and as such we have considered that they decay into gluons and appear as jets at the detector level. To reconstruct the invariant mass of the ALPs from the kinematics of the final state jets, therefore, we have employed the following method. We determine the invariant mass of each pair of jets taking into account all possible combinations and then select those that have an invariant mass closest to the assumed ALP mass. Some representative jet pair invariant mass $M_{jj}$ distributions after BDT cuts determined in this way are displayed in Fig.~\ref{fig:Inv_Mass_Reco}.

\begin{figure}[h]
\centering
\includegraphics[width=7cm, height=5cm]{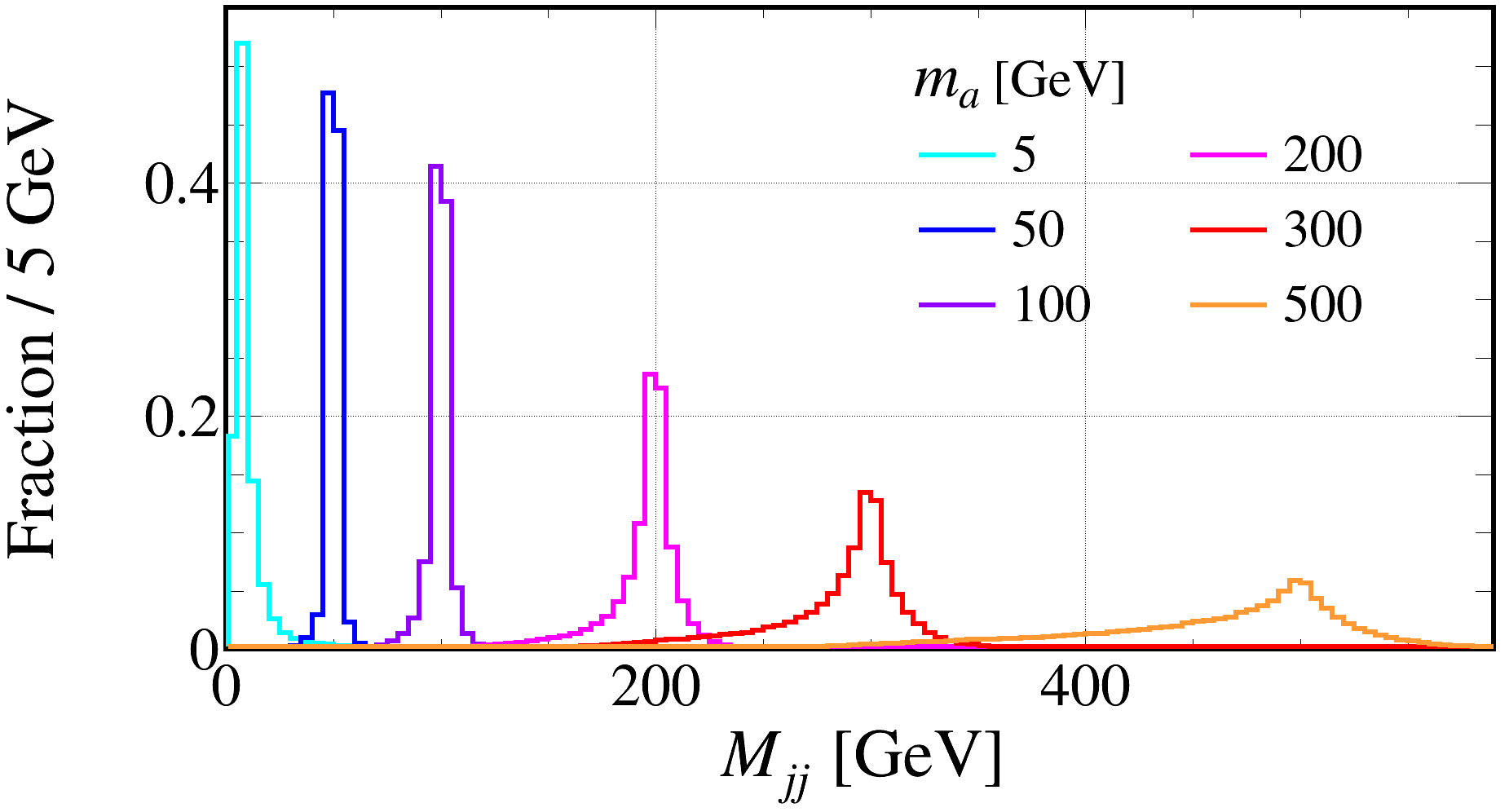}\,\,\,\,
\includegraphics[width=7cm, height=5cm]{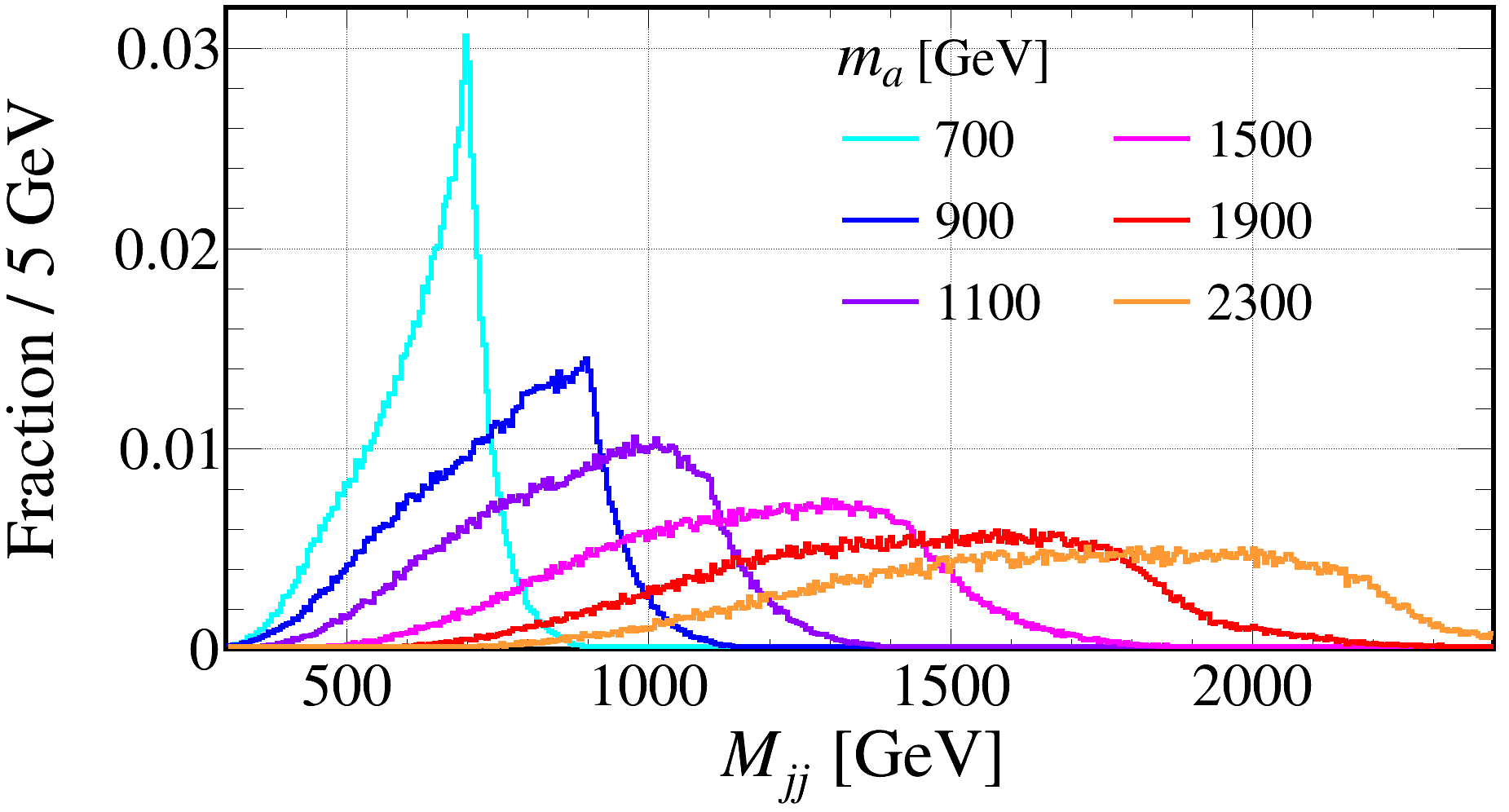}
\caption{
Reconstruction of the invariant mass of ALPs for various representative ALP mass assumptions. The histograms represent distributions of invariant mass constructed from a jet pair in an event whose combined invariant mass is closest to the assumed ALP mass. The parameter $C_{\tilde{G}}/f_a$ is fixed at $10^{-2} \iTeV$ for all signal distributions. 
}
\label{fig:Inv_Mass_Reco}
\end{figure}

As we can see from Fig.~\ref{fig:Inv_Mass_Reco}, for ALP masses up to about 1 TeV a sharp peak for the $M_{jj}$ distribution can be observed around the respective ALP mass. However, for masses above 1 TeV the distribution becomes flatter and the peak appears slightly below the ALP mass.
We generated the signal sample with higher center-of-mass energies, such as the 100 TeV, and found that the $M_{jj}$ can once again be reconstructed with the peak around the ALP mass.
This suggests that the validation of this method depends on the ALP mass and center-of-mass energy, 
and the ALP mass can be reconstructed in this method when $m_a \lesssim$ 1 TeV at the HL-LHC.

\subsection{Effects of Systematic Uncertainties}
\label{subsec:Syst_Unc}

Systematic uncertainties arise during experiments due to a number of factors, including but not limited to jet energy scale and resolution, renormalization and factorization scales, parton shower and hadronization models, PDF variation, jet flavor tagging, etc. 
As shown and calculated from Table~\ref{tab:Crsc} and~\ref{tab:All_Eff}, when $m_a = 500$ GeV, the number of total background events $N_b$ and that of signal events $N_s$ after all cuts are $7.6 \times 10^{12}$ and $5.0 \times 10^{5}$, respectively. Because $N_b \gg N_s$ and $N_b$ has a very huge value, the systematic uncertainty on total background can affect the final limits greatly.
In this section, we estimate and comment on the effects of systematic uncertainties on the limits.
We consider $0.05\%$, $0.1\%$, $0.5\%$, $1\%$, $5\%$, $10\%$ and $15\%$ relative systematic uncertainties on the expected number of events for the total background and discuss their implications on the upper limits for the benchmark case with $m_a$ = 500 GeV.
Systematic uncertainties are introduced as nuisance parameters in estimating the expected median signal significance $\sigma_{\rm sig}$, and therefore we employ the following formula to approximate its value since $N_b \gg N_s$.
\begin{equation}
\sigma_{\rm sig} = \frac{N_s}{\sqrt{N_b + \delta_b^2}}\, ,
\end{equation}
where $N_s$ and $N_b$ are expected numbers of signal and total background events, respectively, after all selection cuts and $\delta_b = r_b \cdot N_b$ denotes uncertainty in the number of total background event when considering a relative systematic uncertainty $r_b$.
Table~\ref{tab:Syst_Unc} summarizes the upper bounds achieved with the afore-mentioned representative relative systematic uncertainties at the 2-, 3- and 5-$\sigma$ confidence levels for the benchmark case with $m_a$ = 500 GeV.

\begin{table}[h]
\centering
\begin{tabular}{cccc}
\hline 
\hline
& 2-$\sigma$ & 3-$\sigma$ & 5-$\sigma$\\ 	
\hline 
0\% & $1.1 \, \mltp \, 10^{-2}$ & $1.3 \, \mltp \, 10^{-2}$ & $1.7 \, \mltp \, 10^{-2}$\\
\hline
0.05\% & $9.1 \, \mltp \, 10^{-2}$ & $1.2 \, \mltp \, 10^{-1}$ & $1.5 \, \mltp \, 10^{-1}$\\
0.1\% & $1.4 \, \mltp \, 10^{-1}$ & $1.7 \, \mltp \, 10^{-1}$ & $2.2 \, \mltp \, 10^{-1}$\\
0.5\% & $3.2 \, \mltp \, 10^{-1}$ & $4.0 \, \mltp \, 10^{-1}$ & $5.1 \, \mltp \, 10^{-1}$\\
   1\% & $4.6 \, \mltp \, 10^{-1}$ & $5.6 \, \mltp \, 10^{-1}$ & $7.3 \, \mltp \, 10^{-1}$\\
  5\% & $1.0$ & $1.3$ & $1.7$\\
10\% & $1.5$ & $1.8$ & $2.3$\\
15\% & $1.8$ & $2.2$ & $2.9$\\		
\hline 
\hline
\end{tabular} 
\caption{Upper limits on the parameter $C_{\tilde{G}}/f_a$ (${\rm TeV^{-1}}$) achieved by considering $0.05\%$, $0.1\%$, $0.5\%$, $1\%$, $5\%$, $10\%$ and $15\%$ relative systematic uncertainties on the number of total background events for the benchmark case with $m_a$ = 500 GeV. The first row represents the limits achieved without any systematic uncertainties on the background. 
}
\label{tab:Syst_Unc}
\end{table}

Since the  expected number of multi-jet events  is much higher than that of other background processes combined, 
the effects of systematic uncertainties on the total number of background events is highly governed by the multi-jet  process.
Hence, we have checked systematic uncertainties on the multi-jet final state from current LHC experimental studies~\cite{ATLAS:2013dpd,ATLAS:2014hvo,ATLAS:2015uwa,ATLAS:2017bje}, and also the projections at the HL-LHC~\cite{HLLHC:sysunc}.
The ATLAS studies at center-of-mass energies of 7 TeV~\cite{ATLAS:2013dpd,ATLAS:2014hvo}, 8 TeV~\cite{ATLAS:2015uwa} and 13 TeV~\cite{ATLAS:2017bje} show that the combined systematic uncertainty of calibrated jets could lie below $\sim$ 2\%.
Ref.~\cite{HLLHC:sysunc} recommends values of relative systematic uncertainties at the HL-LHC for different final state objects. The suggested systematic uncertainties on jets include 0-2\%  due to the pile-up effect, and 0.1-0.2\%, 0.1-0.5\%, 0.75\%  due to measurements of absolute Jet Energy Scale (JES), relative JES, jet flavour, respectively.
Therefore, at future HL-LHC experiments, with upgraded detector design and technology, it is possible to achieve $\lesssim 1\%$ systematic uncertainty on the multi-jet process.

To estimate the effects of different systematic uncertainties on the final upper limits in the considered mass range, 
we consider 0.1\% and 1\% relative systematic uncertainties on the expected number of the total background events, and repeat the analyses for representative masses.
The 2-$\sigma$ limits on the model parameter $C_{\tilde{G}}/f_a$ with 0\% (solid, black), 0.1\% (dashed, blue), and 1\% (dotted, red) relative systematic uncertainties at the HL-LHC with a center-of-mass energy of 14 TeV and an integrated luminosity of 3 $\iab$ are shown in Fig.~\ref{fig:limits_sys_unc}.

\begin{figure}[h]
\centering
\includegraphics[width=10cm, height=6cm]{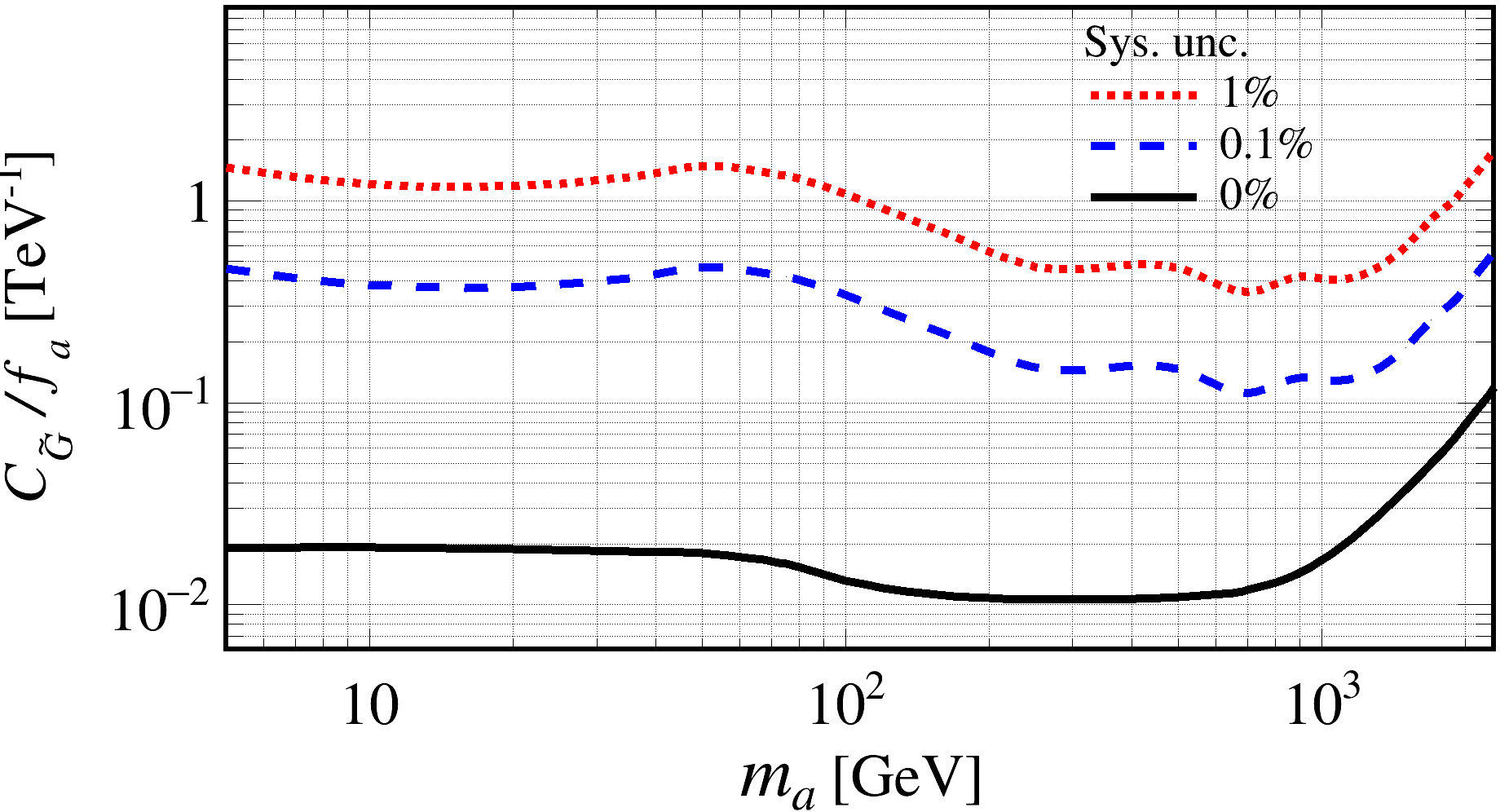}
\caption{
Upper limits on the model parameter $C_{\tilde{G}}/f_a$ at 2-$\sigma$ significance for ALP mass $m_a$ in the range 5-2300 GeV with 1\% (dotted, red), 0.1\% (dashed, blue) and 0\% (solid, black) relative systematic uncertainties on the number of total background events at the HL-LHC with a center-of-mass energy of 14 TeV and an integrated luminosity of 3 $\iab$.
}
\label{fig:limits_sys_unc}
\end{figure}

The results show that the limits are very sensitive to the systematic uncertainty. Even a small relative systematic uncertainty on the total background can weaken the limits. This is mainly because, as we have mentioned above, the number of background events $N_b$ after all cuts is very huge and much larger than the number of signal events $N_s$.
Since the background is dominated by the multi-jet process, to reduce the effects of systematic uncertainty and get strong limits, it is important to reduce the multi-jet background for future experimental analyses.

One method is to apply more stringent BDT cuts. However, as shown in Fig.~\ref{fig:BDT_TBG}, for example, stringent BDT cut is related to the tail distribution of the background. To reduce the statistical fluctuation and  obtain smooth tail distribution, huge number of multi-jet background events need to be generated.
Limited by our computational resources, results in Table~\ref{tab:Syst_Unc} have been achieved by generating $\sim 1.26 \, \mltp \, 10^8$ total background events and $10^6$ signal events for each iteration of ALP mass $m_a$ and the parameter $C_{\tilde{G}}/f_a$ which has amounted to $\sim$ 40 TB of data.
Due to lack of sufficient multi-jet background sample, the tail distribution of the background has very big fluctuations. To get reliable results, our analysis has been done by preserving sizable multi-jet background events and, hence, we are not able to reject this background further.
Therefore, the results listed in Table~\ref{tab:Syst_Unc} and Fig.~\ref{fig:limits_sys_unc} merely represent how systematic uncertainties affect the upper limits for the data we have accumulated for this study.
Future experimental analyses with ample computational resources could achieve better results by obtaining more background, especially the multi-jet event samples, and simulating the tail part of the total background much better.
One can also input more observables to the multivariate analysis to reject the multi-jet background, which needs to check more kinematic observables and find those having clear differences between the signal and background, and we leave it for future studies.

\subsection{Validation of the Effective Lagrangian}
\label{subsec:Val_EFT}

\begin{figure}[h]
\centering
\includegraphics[width=7cm, height=4.9cm]{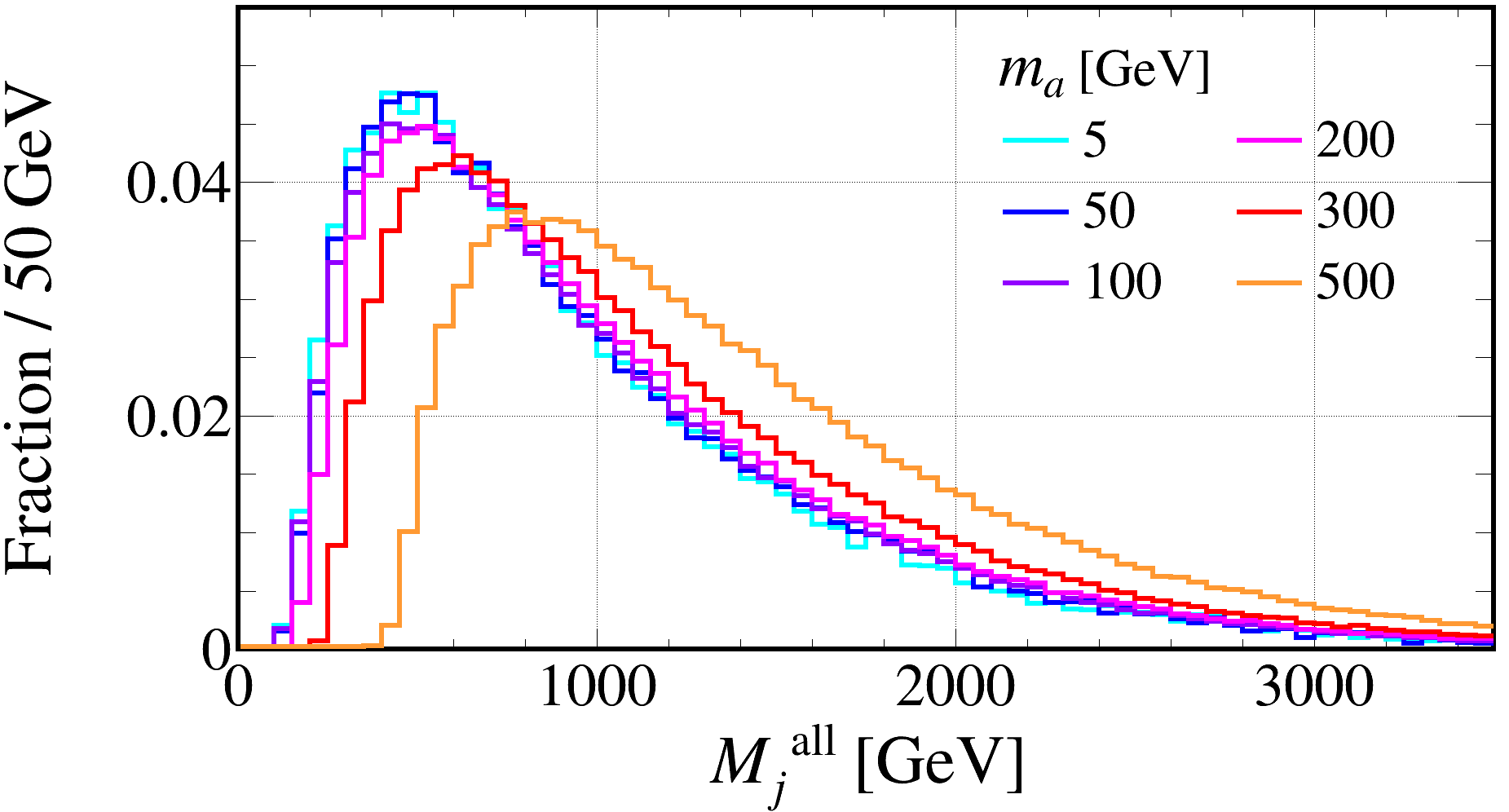}\,\,\,\,
\includegraphics[width=7cm, height=5cm]{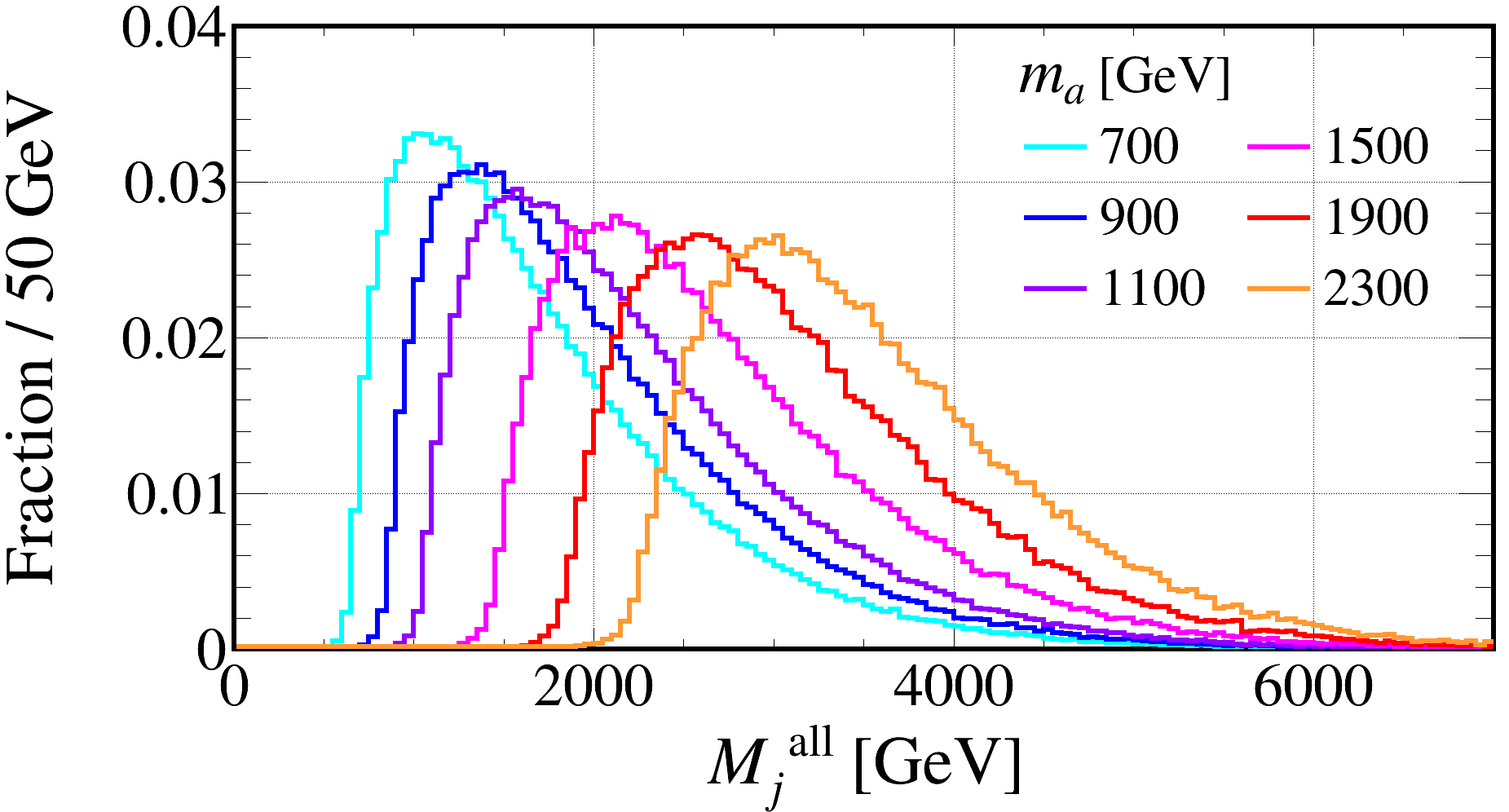}  
\caption{
Distributions of the invariant mass of all final state jets $M_{j}^{\rm all}$ for the signal process with representative ALP masses and $C_{\tilde{G}}/f_a = 10^{-2} \,\, {\rm TeV^{-1}}$ after pre-selection cuts.
}
\label{fig:M_all_val}
\end{figure}

The effective operators are weighted down by the energy scale $f_a$ in the effective Lagrangian provided in Eq.~(\ref{eqn:Eff_Lag}). This suppression energy scale must be considerably larger than the typical energy scale of the process under investigation in order for the description of the Effective Field Theory (EFT) to be valid. 
That is, it is required that for every generated signal event, the center-of-mass energy of the signal process should be less than the energy scale, i.e. $\sqrt{\hat{s}} \, < \, f_a$~\cite{Brivio:2017ije}. 
Since $\sqrt{\hat{s}}$ cannot be measured directly in experiment and the signal final state consists of all jets, we, approximately, consider the invariant mass of all final state jets $M_{j}^{\rm all}$ as $\sqrt{\hat{s}}$.
Fig.~\ref{fig:M_all_val} shows $M_{j}^{\rm all}$ distributions of the signal with several representative ALP masses after pre-selection cuts.
Considering the validation of the EFT framework, the constraint $M_{j}^{\rm all} < f_a$ needs to be imposed. 
When $f_a$ is 2 TeV, for example, 
a fraction of signal events are invalid for $m_a > 500$ GeV, and thus final limits are weakened for heavy masses. 
However, such effects on limits are small when $f_a$ is very large (for example, 5 TeV) even for heavy masses.
Therefore, the results depend strongly on the symmetry breaking scale $f_a$, especially for the heavy masses.

\section{Conclusion}
\label{sec:Conc}

A wide area of the parameter space spanned by the ALP mass $m_a$ and its couplings to gluons $C_{\tilde{G}}/f_a$ is probed at the HL-LHC with a center-of-mass energy $\sqrt{s}$ = 14 TeV and integrated luminosity $\mathcal{L}_{\rm int}$ = 3 $\iab$. 
Assuming ALPs couple to gluons only, they are produced in association with one jet, $p \, p \,\rightarrow a \, j$, and decay into gluon pairs. Thus, the signal final state has at least three jets.
Seven dominant background processes relevant to our signal are considered and simulated at the detector level. 
Signal processes are scanned for model parameters of $m_a$ and $C_{\tilde{G}}/f_a$ and events are generated for each specific combination.
To consider the long-lived properties of ALPs, a specialized Delphes module~\cite{Delphes:wDisp} has been employed to deal with ALPs decaying into displaced jets, leading to corresponding observables such as the transverse and longitudinal displacements of one jet $V_T(j)$ and $V_z(j)$.

For data analyses, we first apply pre-selection cuts to select the signal final state.
Owing to large expected number of background events,
thirty observables including $V_T(j)$ and $V_z(j)$ are input to the TMVA package to perform multivariate analyses via the BDT algorithm.
We present the distributions of representative observables and the corresponding BDT responses for the signal process with benchmark mass $m_a = 500$ GeV and the background process in Fig.~\ref{fig:Input_Disc_Var} and Fig.~\ref{fig:BDT_TBG}, respectively, and
BDT response distributions for various $m_a$ assumptions are depicted in Fig.~\ref{fig:BDT_Dis}.
The highest ranked observables are given in different mass ranges, according to their importance when discriminating between the signal and background, which are demonstrated in Fig.~\ref{fig:Input_Disc_Var_masses}.

The BDT responses are exploited to discriminate signal and background events and achieve the best sensitivities.
We list selection efficiencies for the signal processes with representative ALP masses and fixed $C_{\tilde{G}}/f_a = 10^{-2} \,\, {\rm TeV^{-1}}$ and the background processes in Table~\ref{tab:All_Eff}, and 
upper limits on the model parameter $C_{\tilde{G}}/f_a$ for ALP mass $m_a$ in the range 5$-$2300 GeV at 2-, 3- and 5-$\sigma$ significances at the HL-LHC with 3 $\iab$ integrated luminosity are shown in Fig.~\ref{fig:Upper_Limit}.
Our results show that the best upper limits at 2-, 3- and 5-$\sigma$ significances are $1.09 \times 10^{-2}$, $1.33 \times 10^{-2}$ and $1.71 \times 10^{-2} \,\, {\rm TeV^{-1}}$ for $m_a \sim 500$ GeV, respectively. 

To reconstruct the invariant mass of the ALPs from the kinematics of the final state jets, we plot distributions of invariant mass constructed from a jet pair in an event whose combined invariant mass is closest to the assumed ALP mass and find that the ALP mass can be reconstructed in this method when $m_a \lesssim$ 1 TeV at the HL-LHC.

To estimate the effects of systematic uncertainties, we consider $0.05\%$, $0.1\%$, $0.5\%$, $1\%$, $5\%$, $10\%$ and $15\%$ relative systematic uncertainties on the expected number of events for the total background and check their implications on the upper limits for the benchmark case with $m_a$ = 500 GeV.
We also consider 0.1\% and 1\% relative systematic uncertainties, and repeat the analyses for representative masses.
The 2-$\sigma$ limits on the model parameter $C_{\tilde{G}}/f_a$ in the considered mass range are shown in Fig.~\ref{fig:limits_sys_unc}.
Results show that upper limits are very sensitive to the systematic uncertainty for the data we have accumulated for this study. This is mainly because the number of background events after all cuts is still very huge and much larger than the number of signal events. Since the background is dominated by the multi-jet process, to reduce the effects of systematic uncertainty, it is important to reduce the multi-jet background for future experimental analyses. The possible methods are discussed.

To check the validation of the EFT framework, we, approximately, consider the invariant mass of all final state jets $M_{j}^{\rm all}$ as the center-of-mass energy of the signal process $\sqrt{\hat{s}}$ and plot its distributions with several representative ALP masses after pre-selection cuts. We find that the limits depend strongly on the symmetry breaking scale $f_a$, especially for the heavy masses.

Granted that there is room for improvement, the analyses done in this study and the sensitivities achieved shed light 
on vast previously unprobed regions of heavy ALP physics through the ALP coupling to gluons at the HL-LHC and give an insight into the sensitivities that can be achieved at future upgraded proton-proton colliders.

\acknowledgments

We thank Lingxiao Bai, Haiyong Gu, Ying-nan Mao, Diego Redigolo and Minglun Tian for useful discussions.
F.A. and K.W. are supported by the National Natural Science Foundation of China under grant no.~11905162, the Excellent Young Talents Program of the Wuhan University of Technology under grant no.~40122102, and the research program of the Wuhan University of Technology under grant no.~2020IB024.
Z.S.W. is supported by the Ministry of Science and Technology (MoST) of Taiwan with grant numbers MoST-109-2811-M-007-509 and MoST-110-2811-M-007-542-MY3.
The simulation and analysis work of this paper was completed with the computational cluster provided by the Theoretical Physics Group at the Department of Physics, School of Sciences, Wuhan University of Technology.

\appendix

\newpage
\section{Distributions of Representative Observables}
\label{append:Disc_Var}

\begin{figure}[h]
\centering
\includegraphics[width=4.5cm,height=3.5cm]{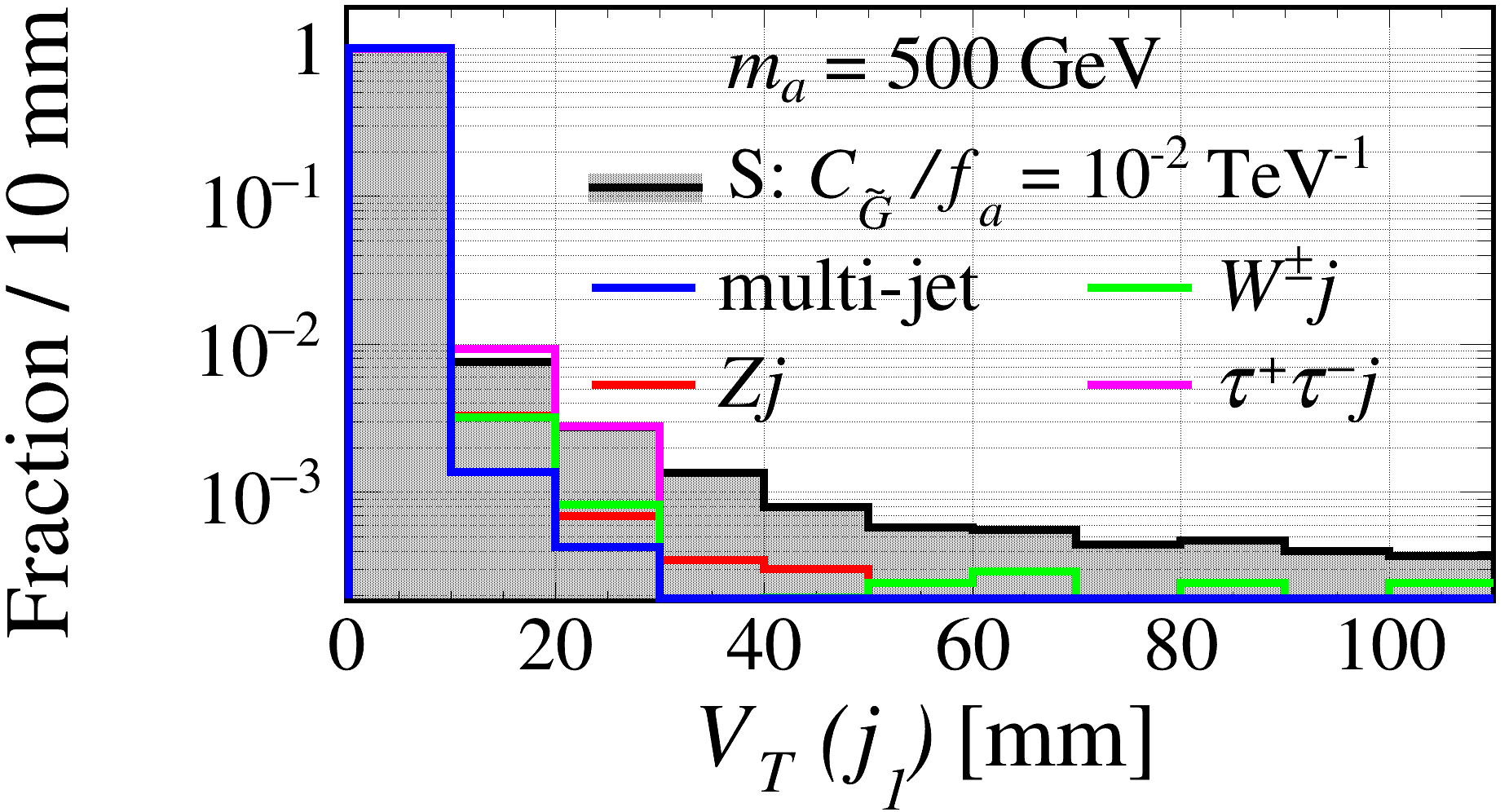}\,\,\,\,\,\,
\includegraphics[width=4.5cm,height=3.5cm]{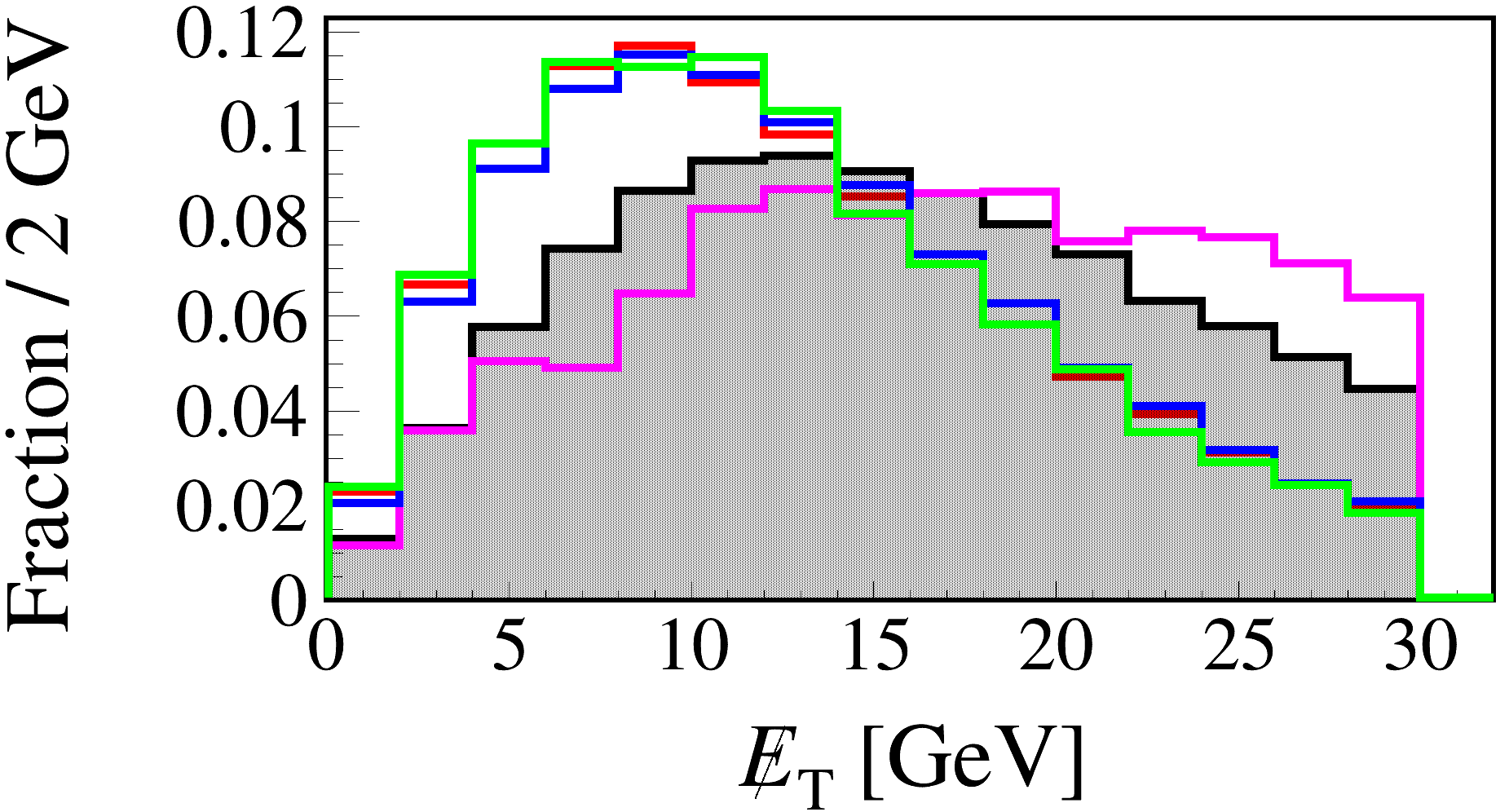}\,\,\,\,\,\,
\includegraphics[width=4.5cm,height=3.5cm]{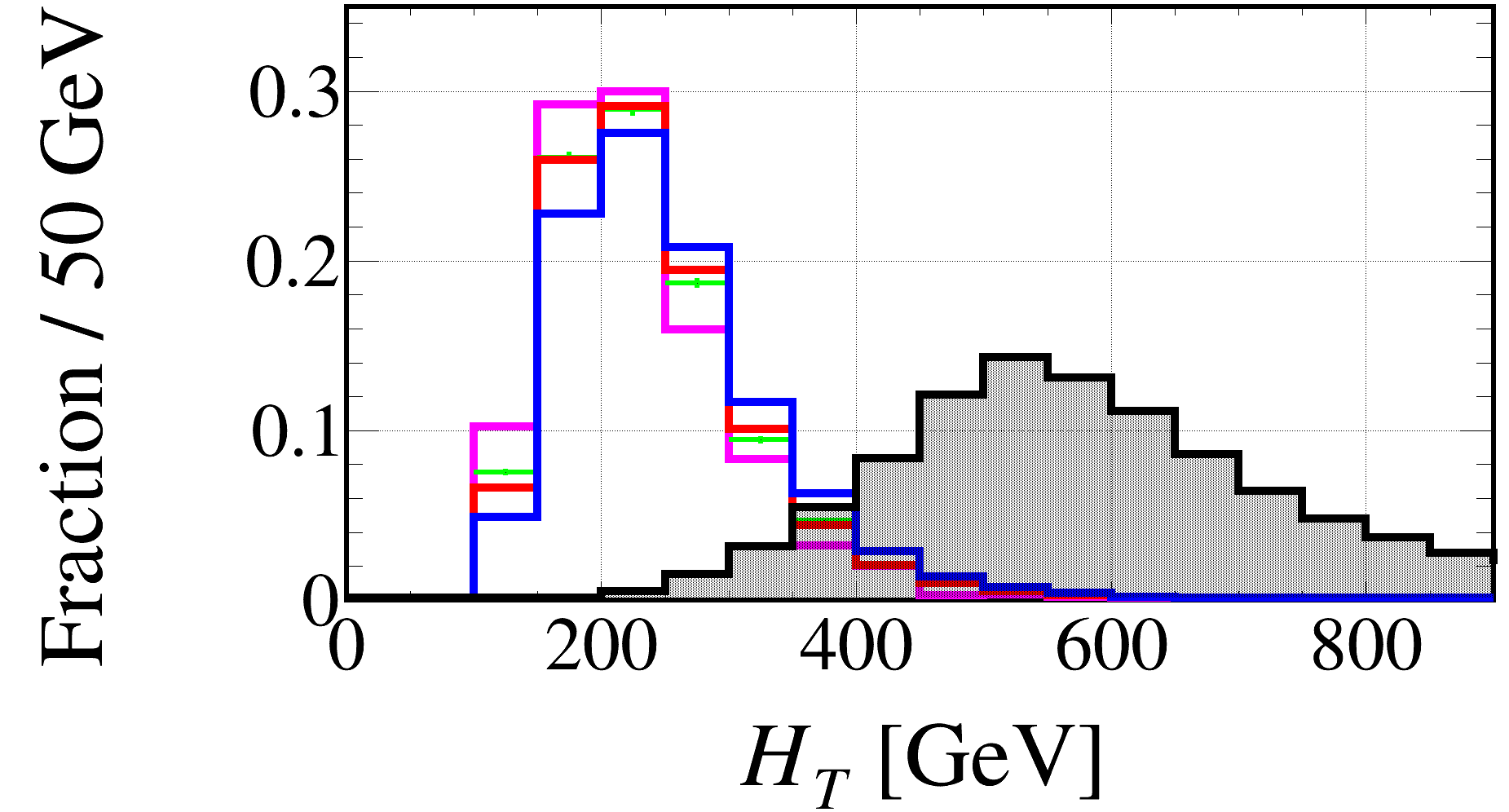}
\par\medskip
\includegraphics[width=4.5cm,height=3.5cm]{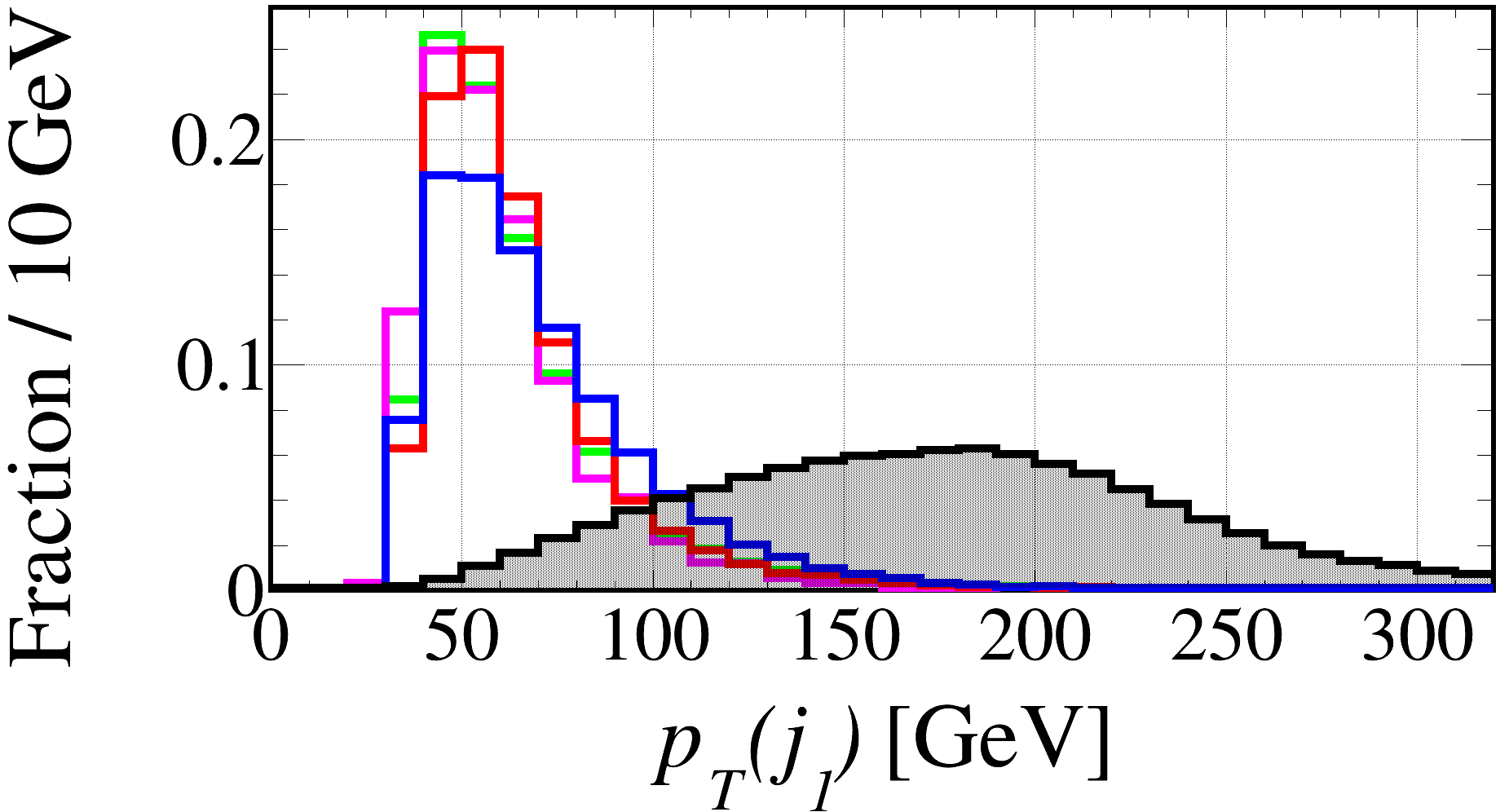}\,\,\,\,\,\,
\includegraphics[width=4.5cm,height=3.5cm]{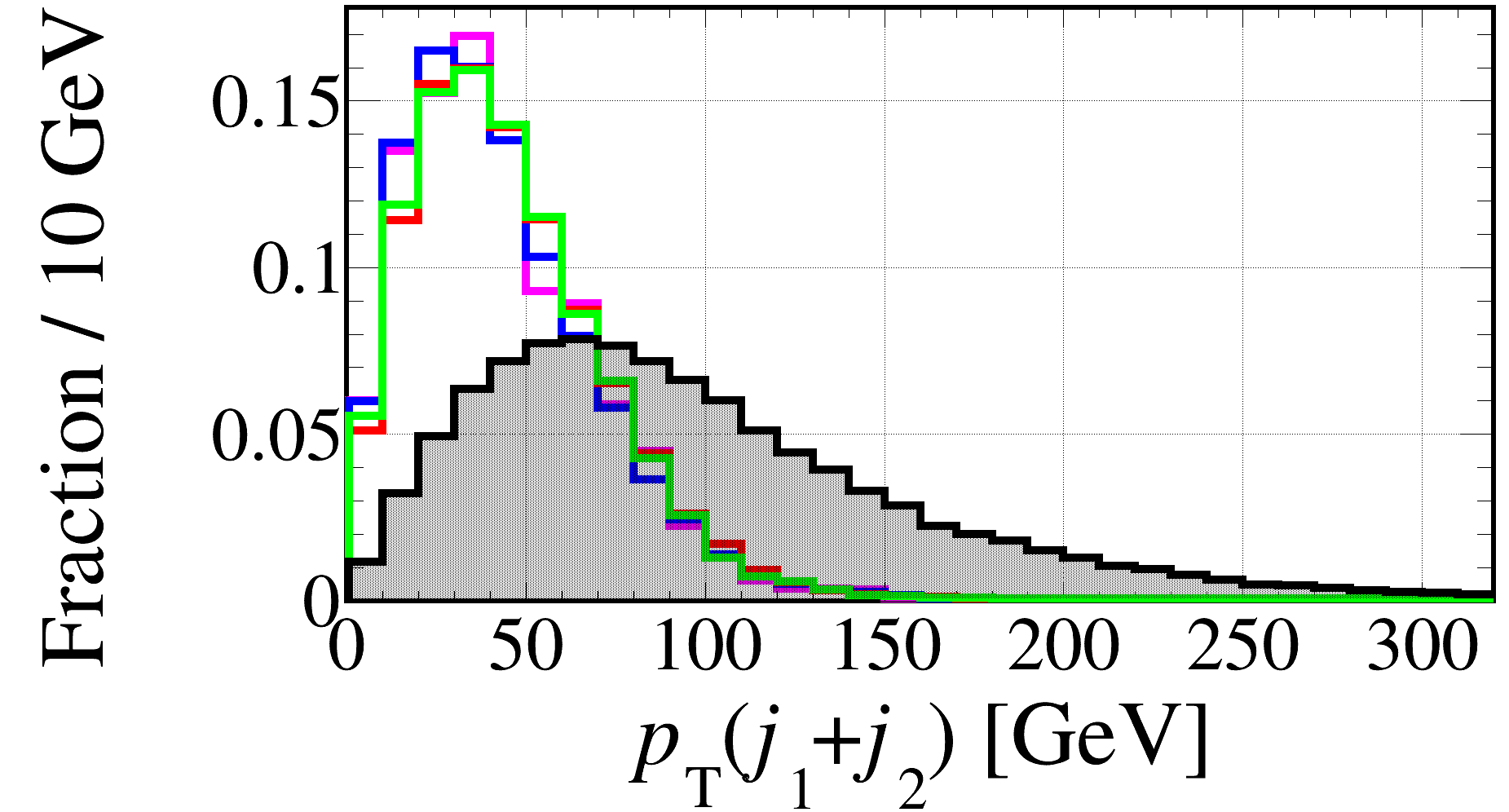}\,\,\,\,\,\,
\includegraphics[width=4.5cm,height=3.5cm]{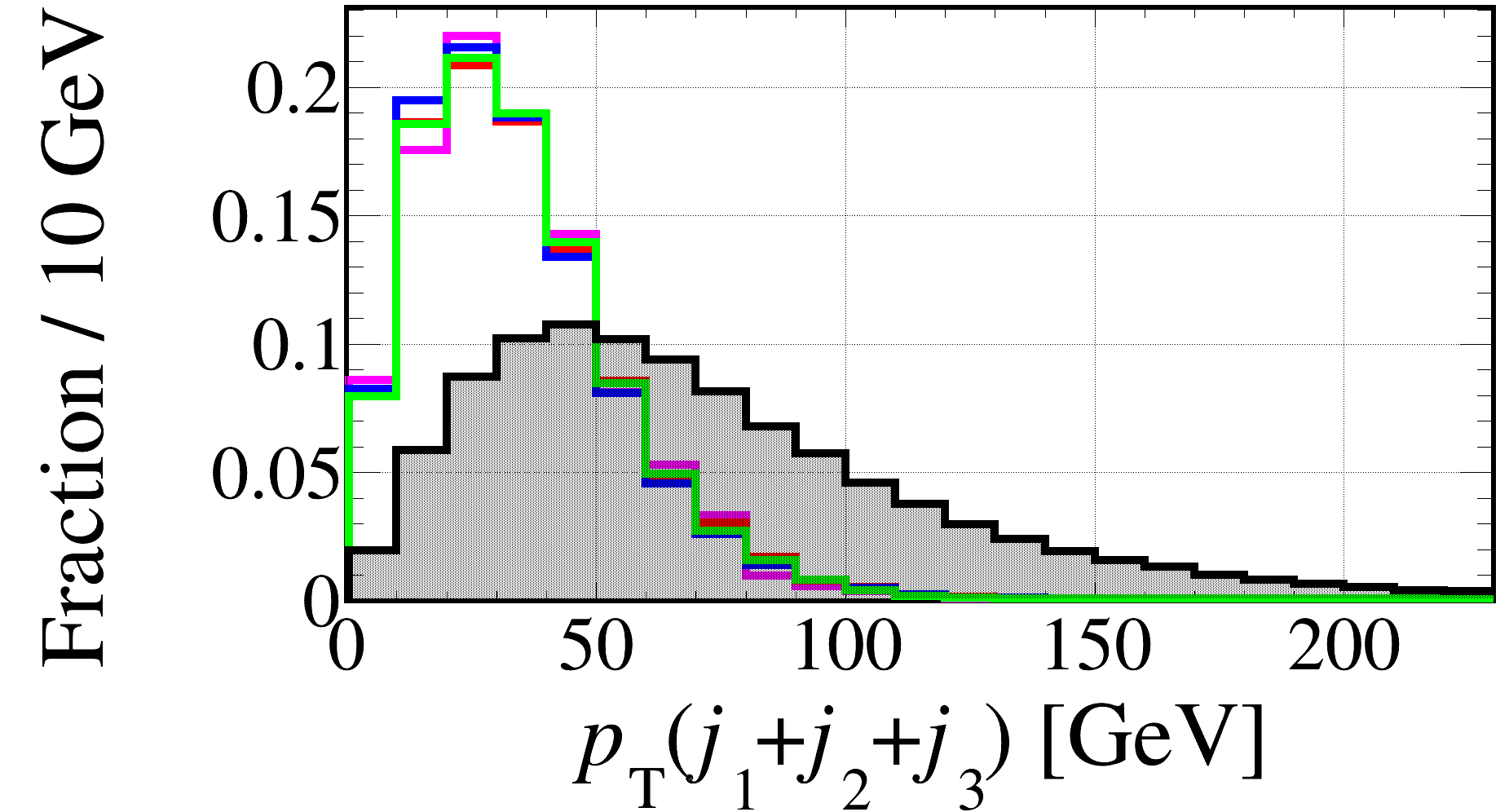}
\par\medskip
\includegraphics[width=4.5cm,height=3.5cm]{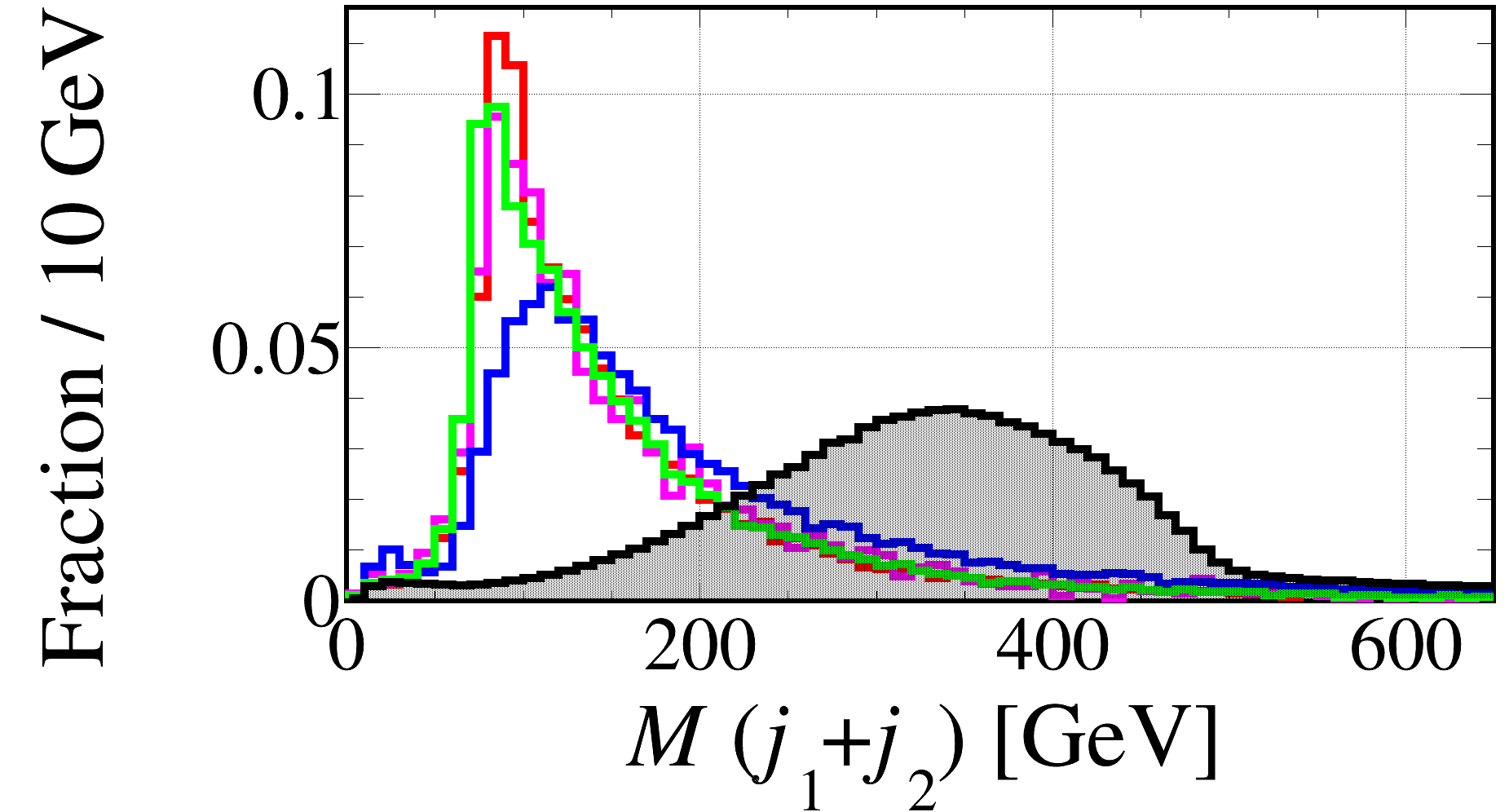}\,\,\,\,\,\,
\includegraphics[width=4.5cm,height=3.5cm]{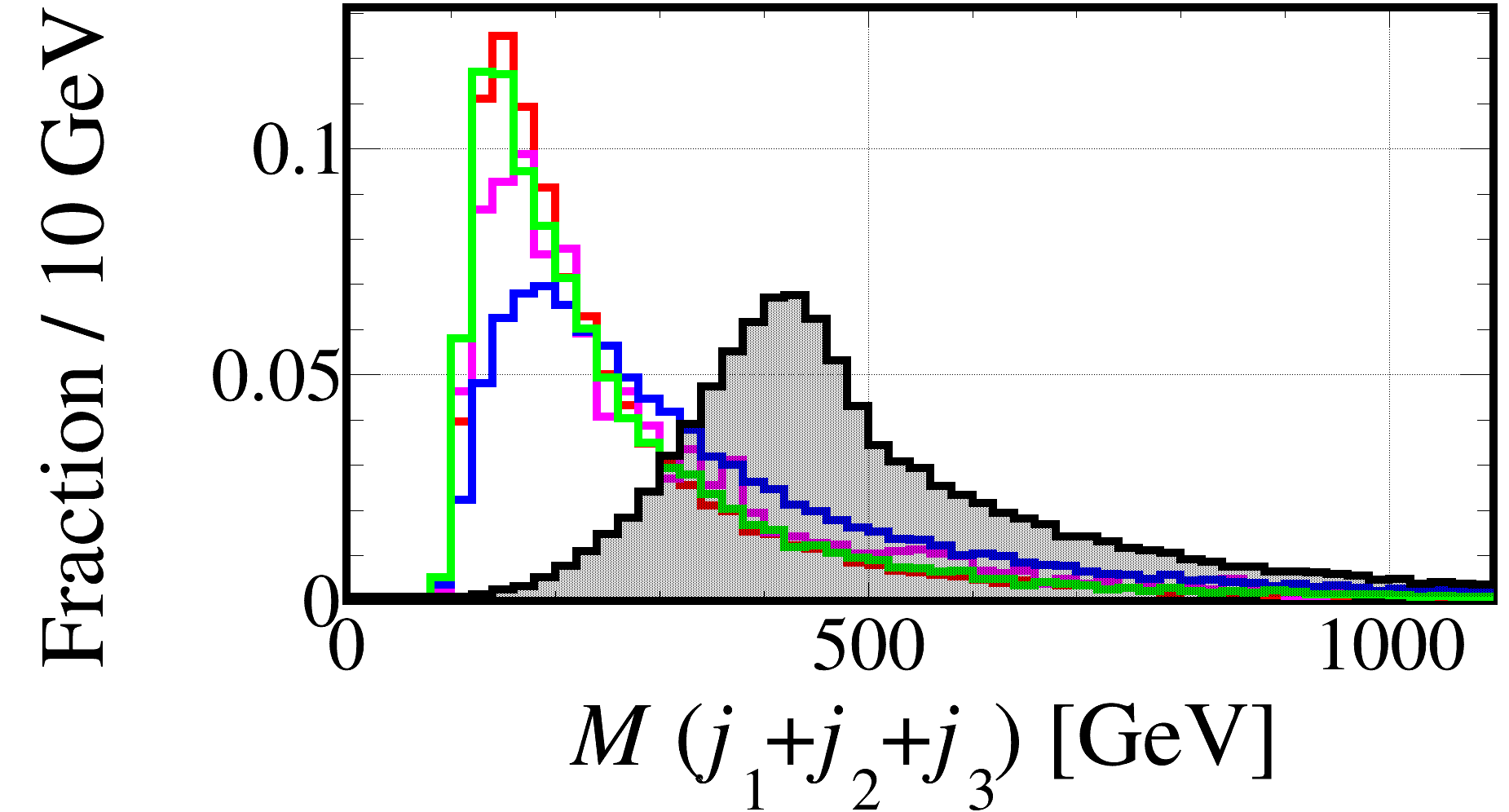}\,\,\,\,\,\,
\includegraphics[width=4.5cm,height=3.5cm]{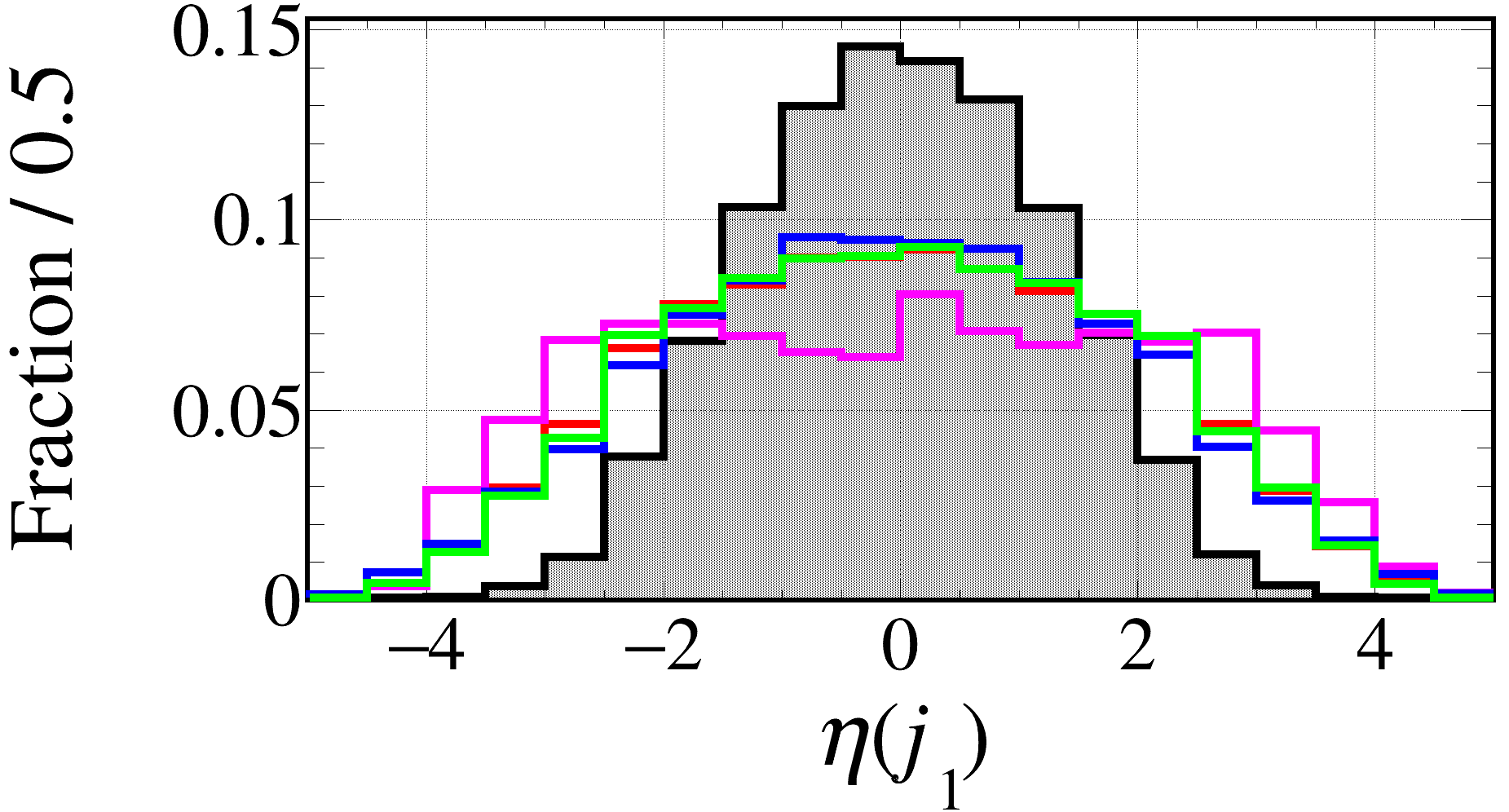}
\caption{
Distributions of representative observables used as input to the multivariate analysis for the benchmark mass $m_{a}$ = 500 GeV and parameter $C_{\tilde{G}}/f_a = 10^{-2}\,\, {\rm TeV}^{-1}$ at the HL-LHC.
}
\label{fig:Input_Disc_Var}
\end{figure}

\newpage
\section{Distributions of BDT Responses}
\label{append:BDT_Dis}

\begin{figure}[h]
\includegraphics[width=4.5cm,height=3.5cm]{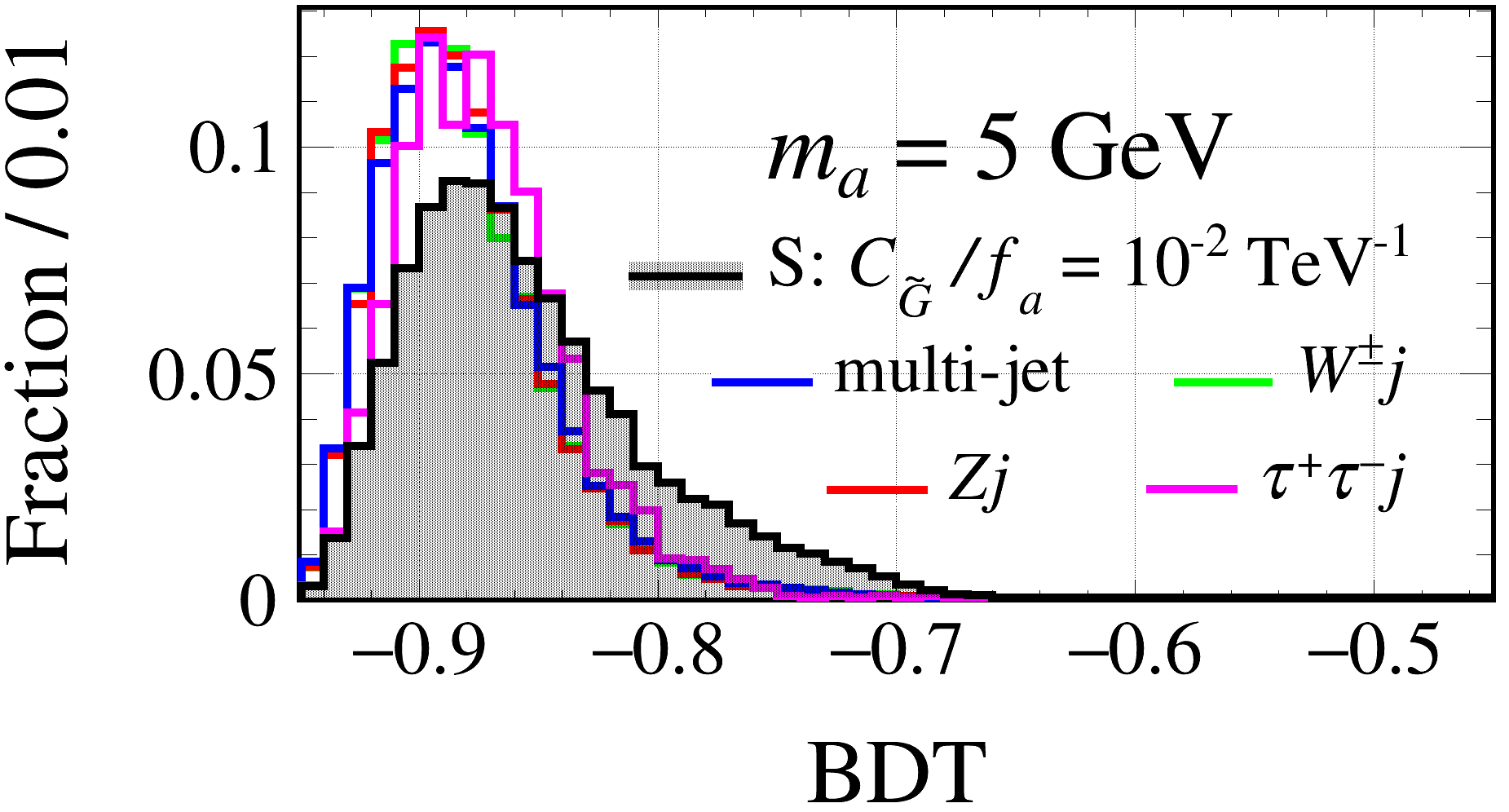}\,\,\,\,\,\,
\includegraphics[width=4.5cm,height=3.5cm]{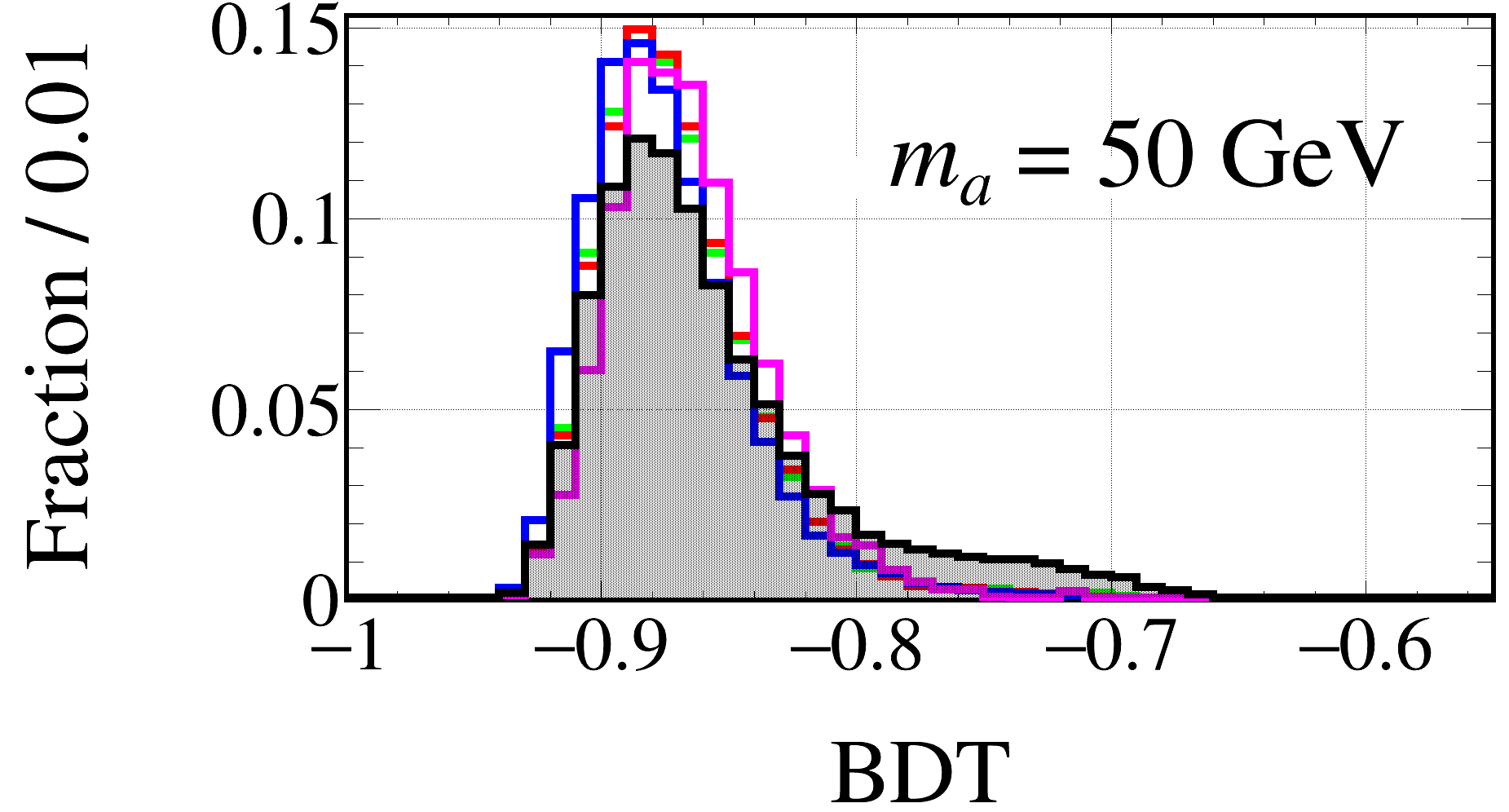}\,\,\,\,\,\,
\includegraphics[width=4.5cm,height=3.5cm]{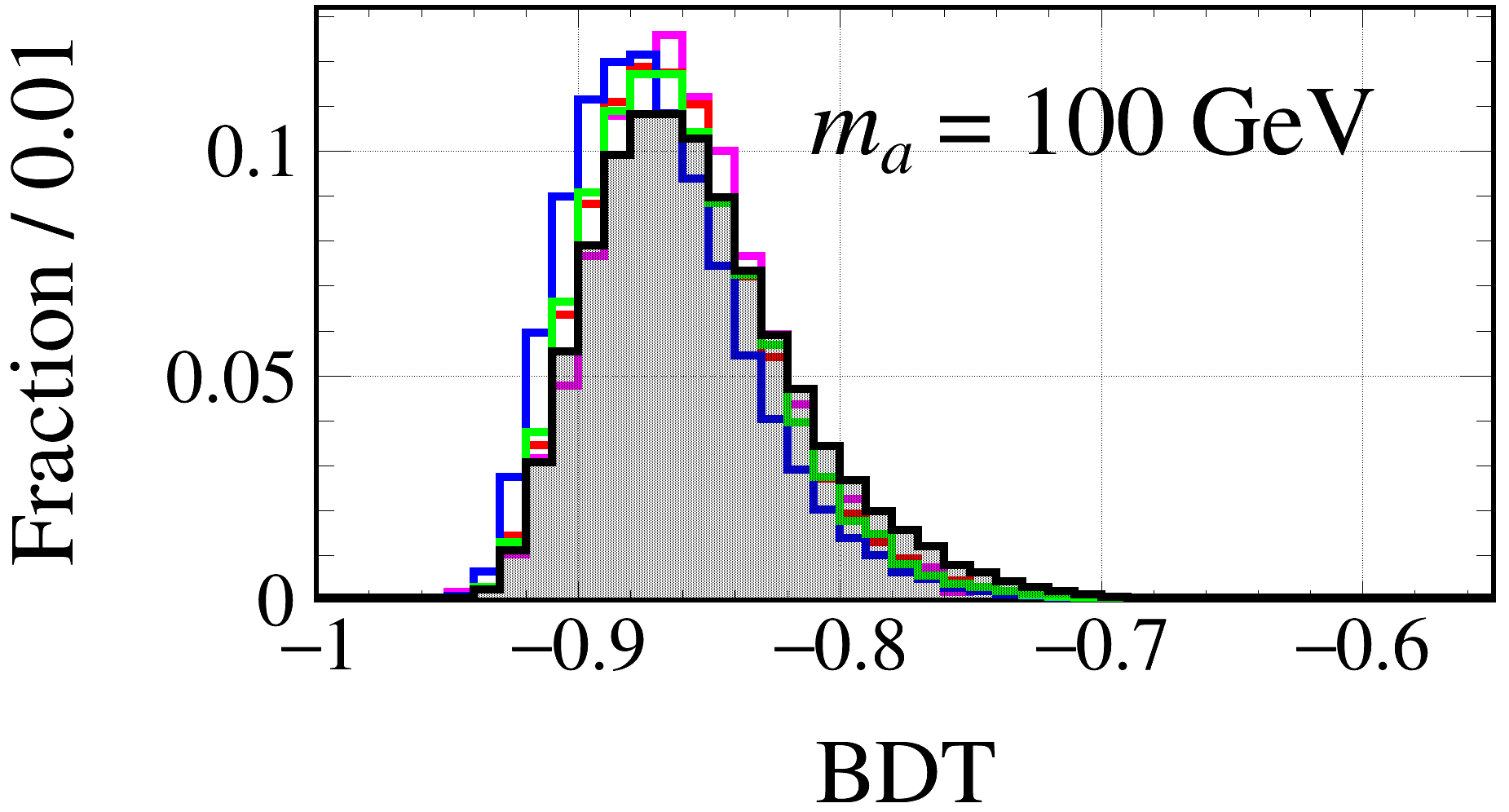}
\par\medskip	
\includegraphics[width=4.5cm,height=3.5cm]{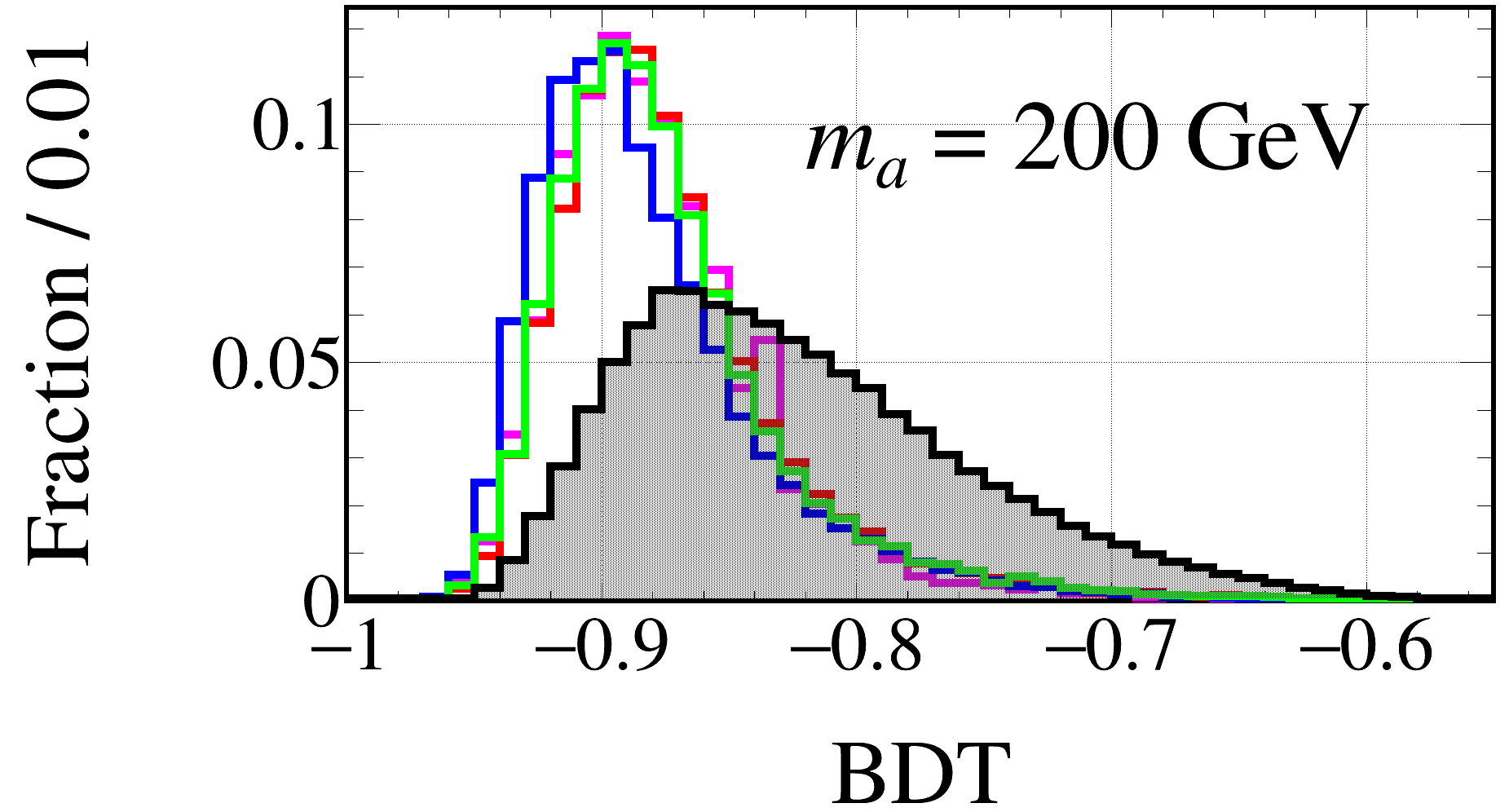}\,\,\,\,\,\,
\includegraphics[width=4.5cm,height=3.5cm]{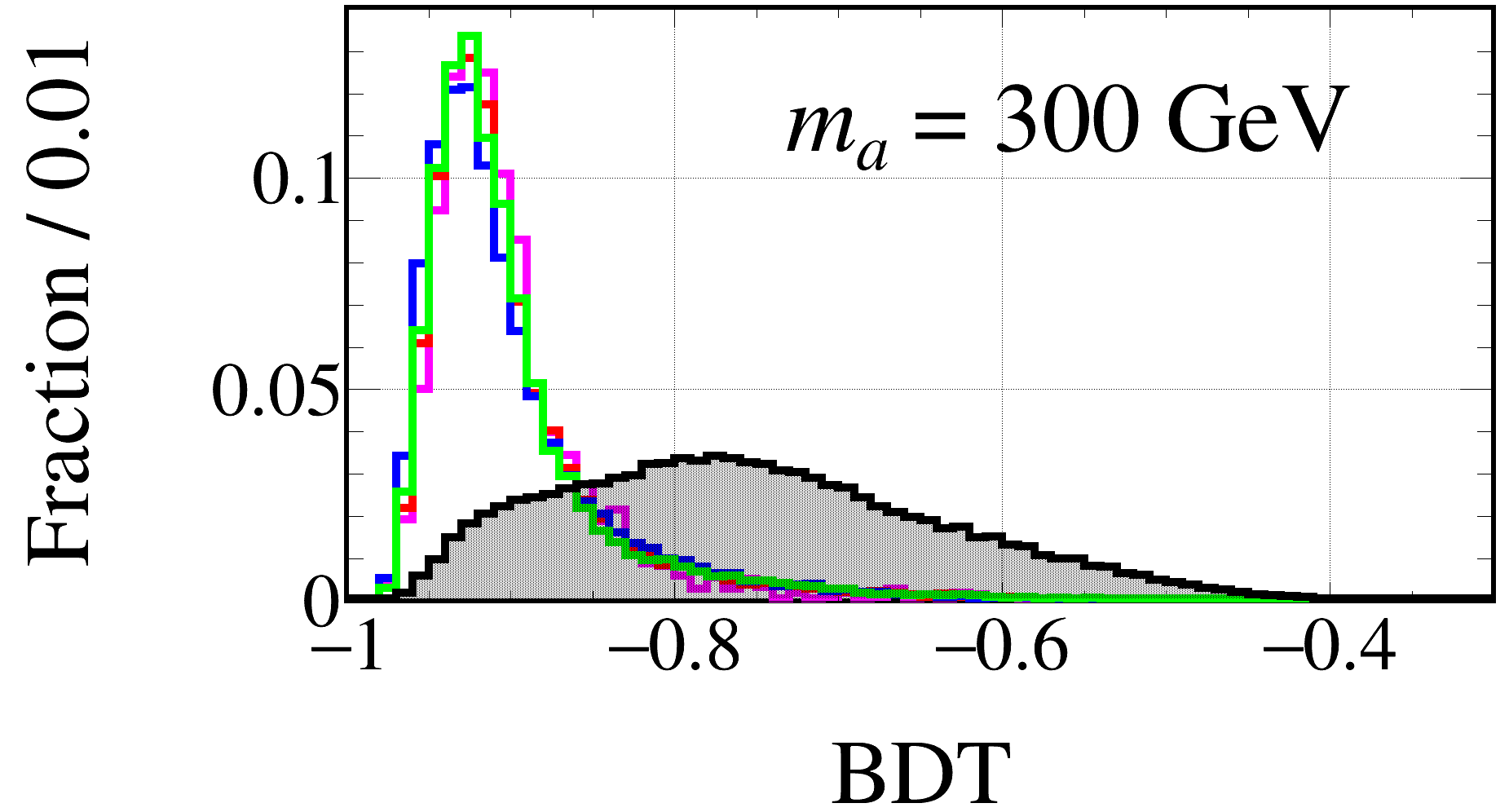}\,\,\,\,\,\,\includegraphics[width=4.5cm,height=3.5cm]{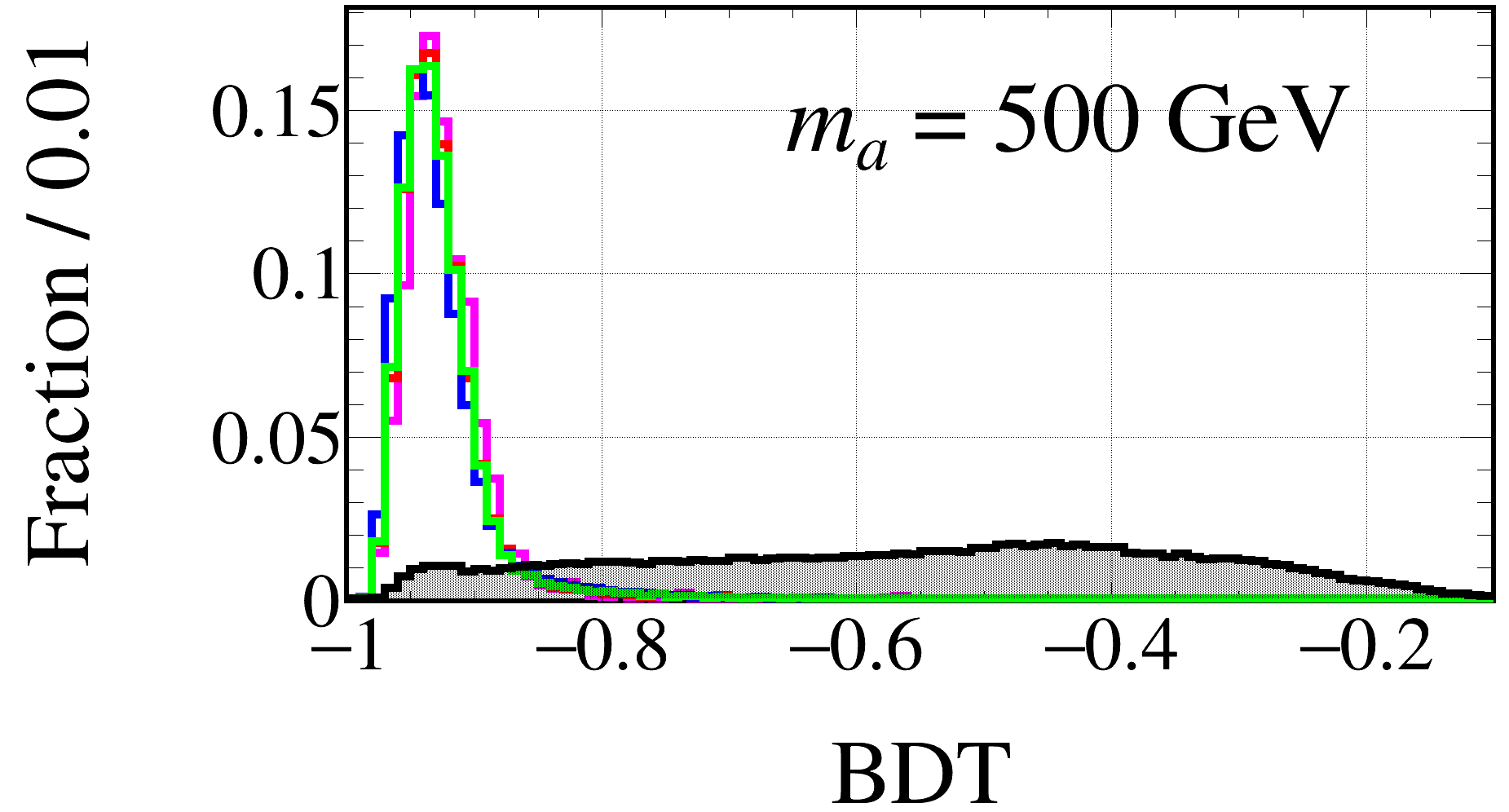}
\par\medskip
\includegraphics[width=4.5cm,height=3.5cm]{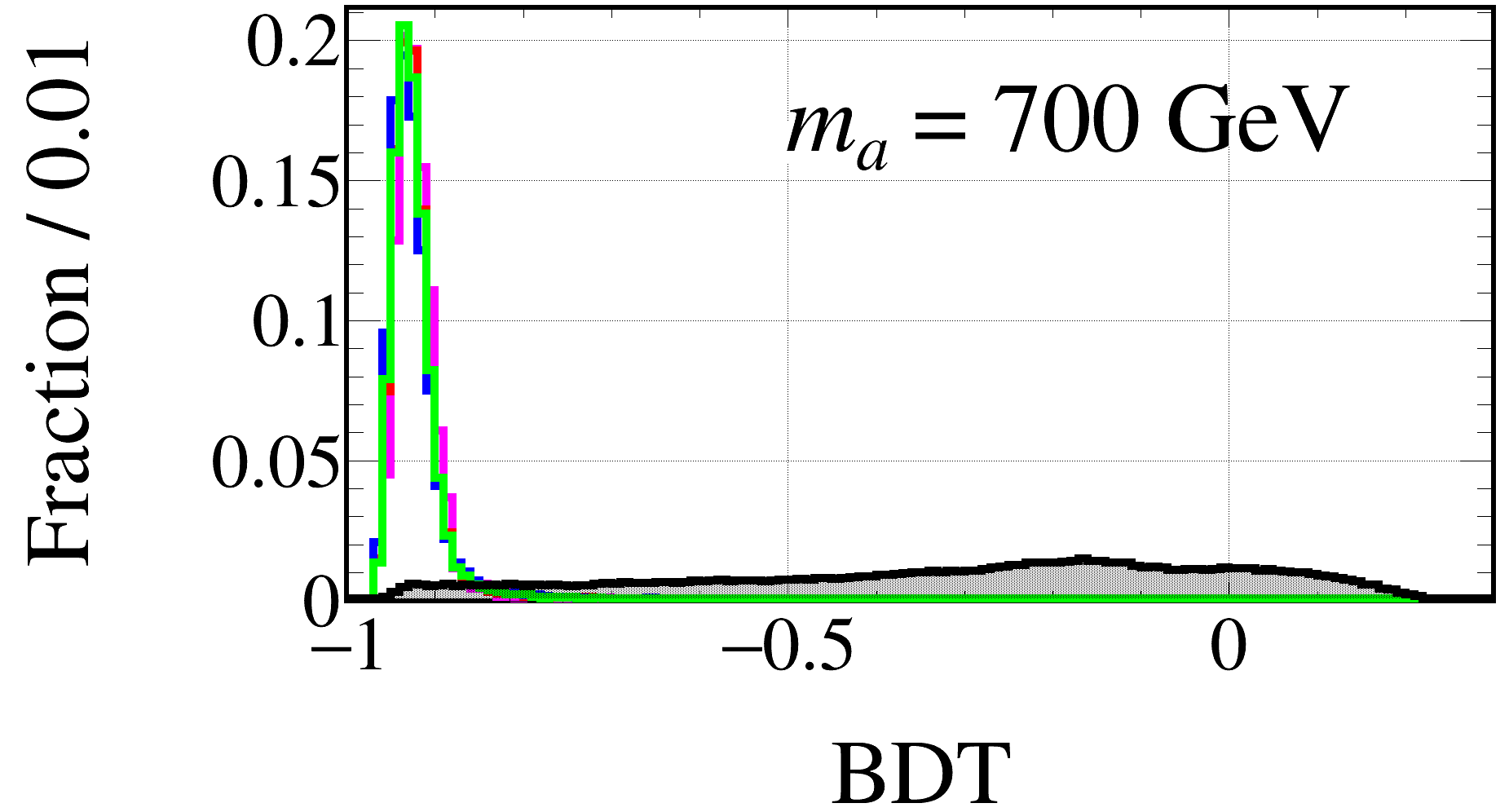}\,\,\,\,\,\,	\includegraphics[width=4.5cm,height=3.5cm]{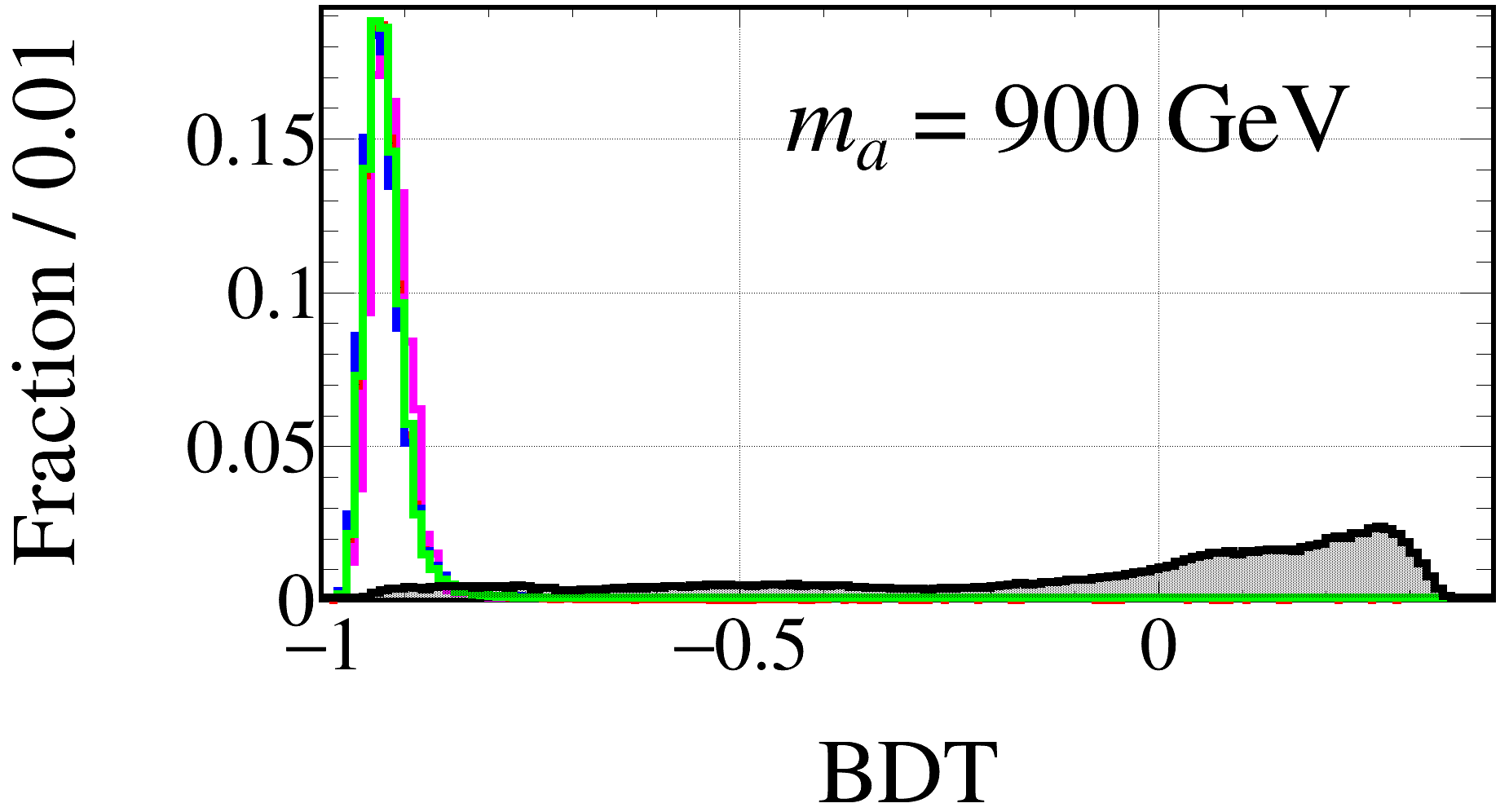}\,\,\,\,\,\,
\includegraphics[width=4.5cm,height=3.5cm]{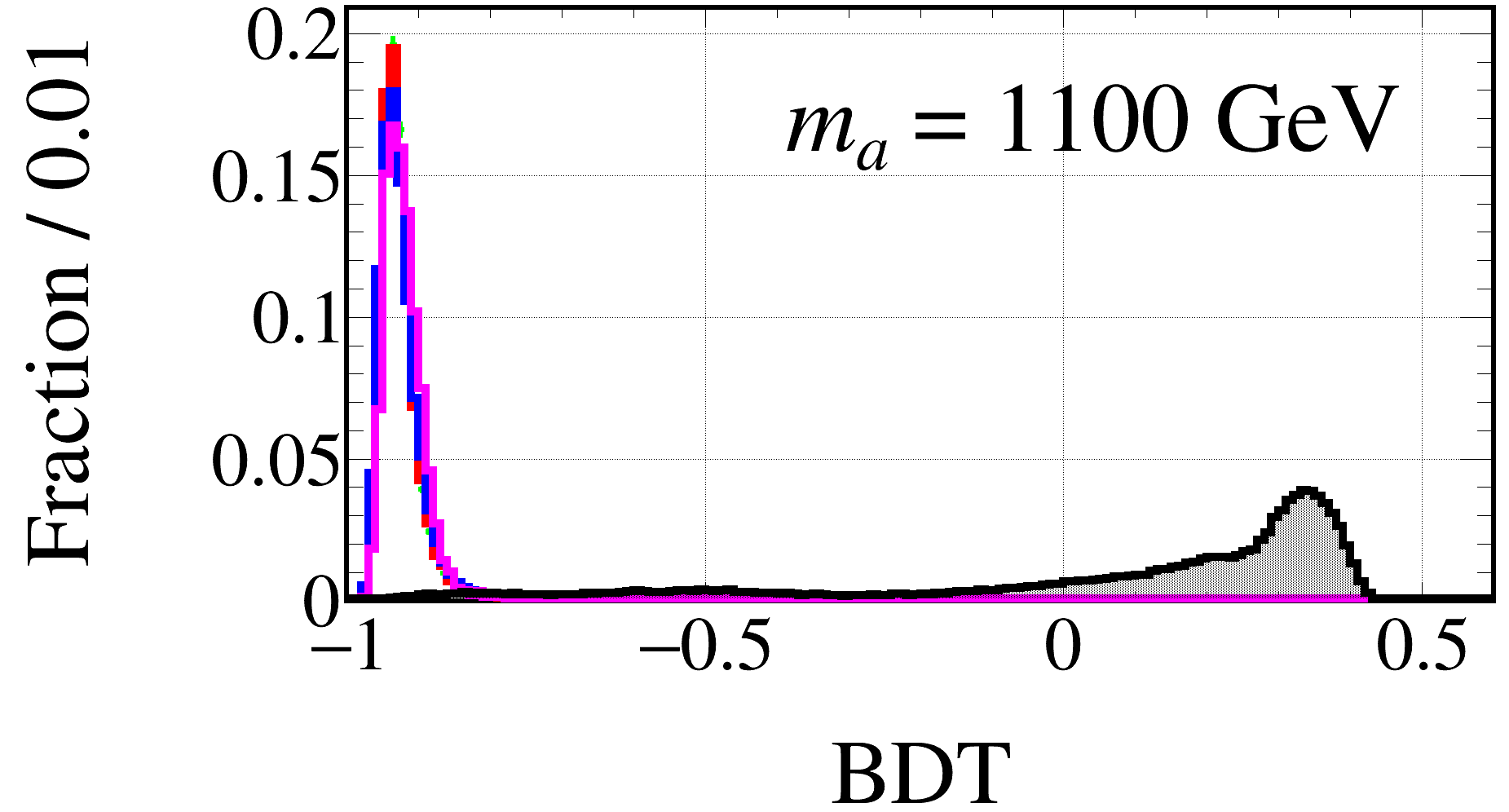}
\par\medskip
\includegraphics[width=4.5cm,height=3.5cm]{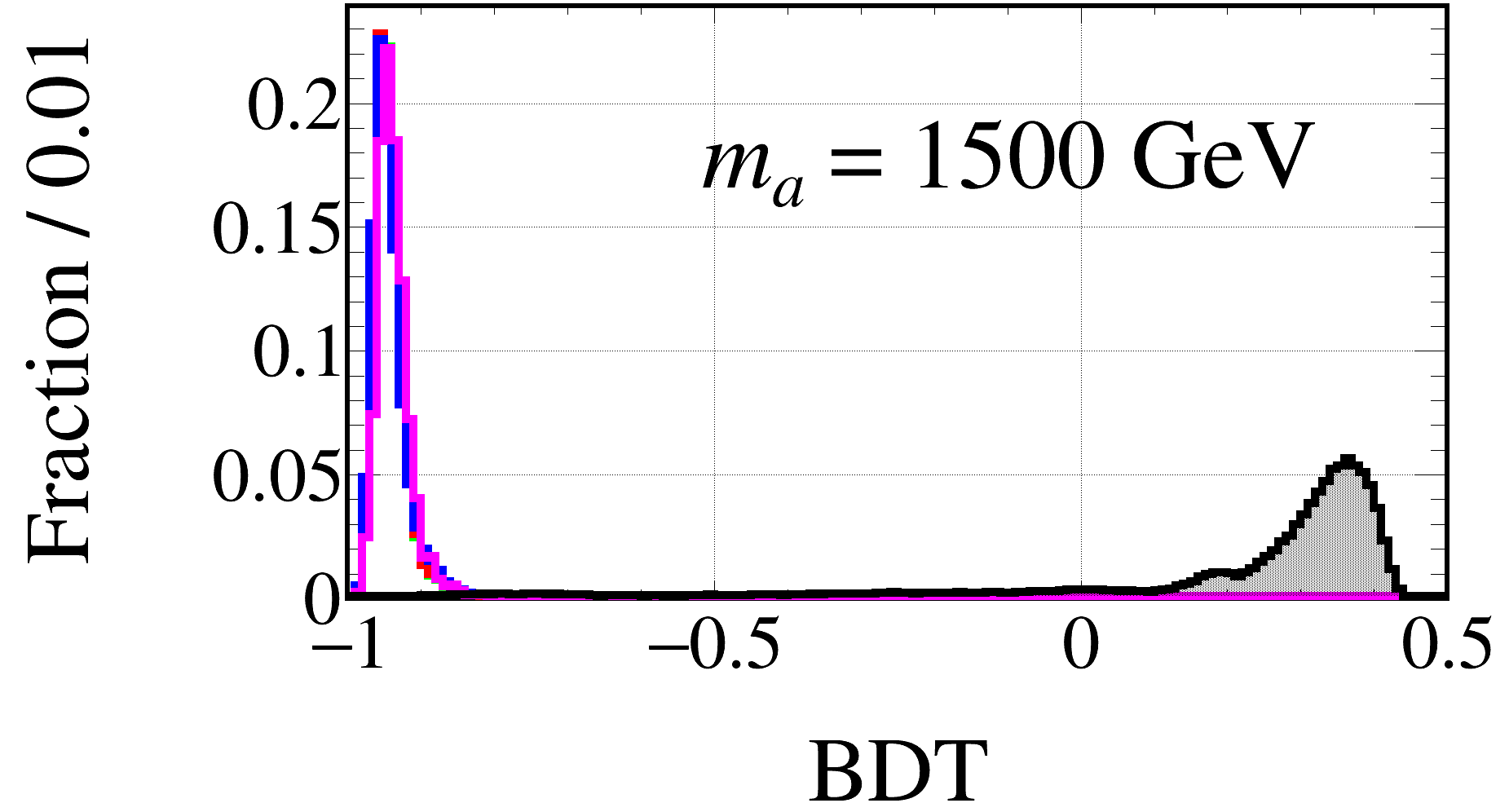}\,\,\,\,\,\,
\includegraphics[width=4.5cm,height=3.5cm]{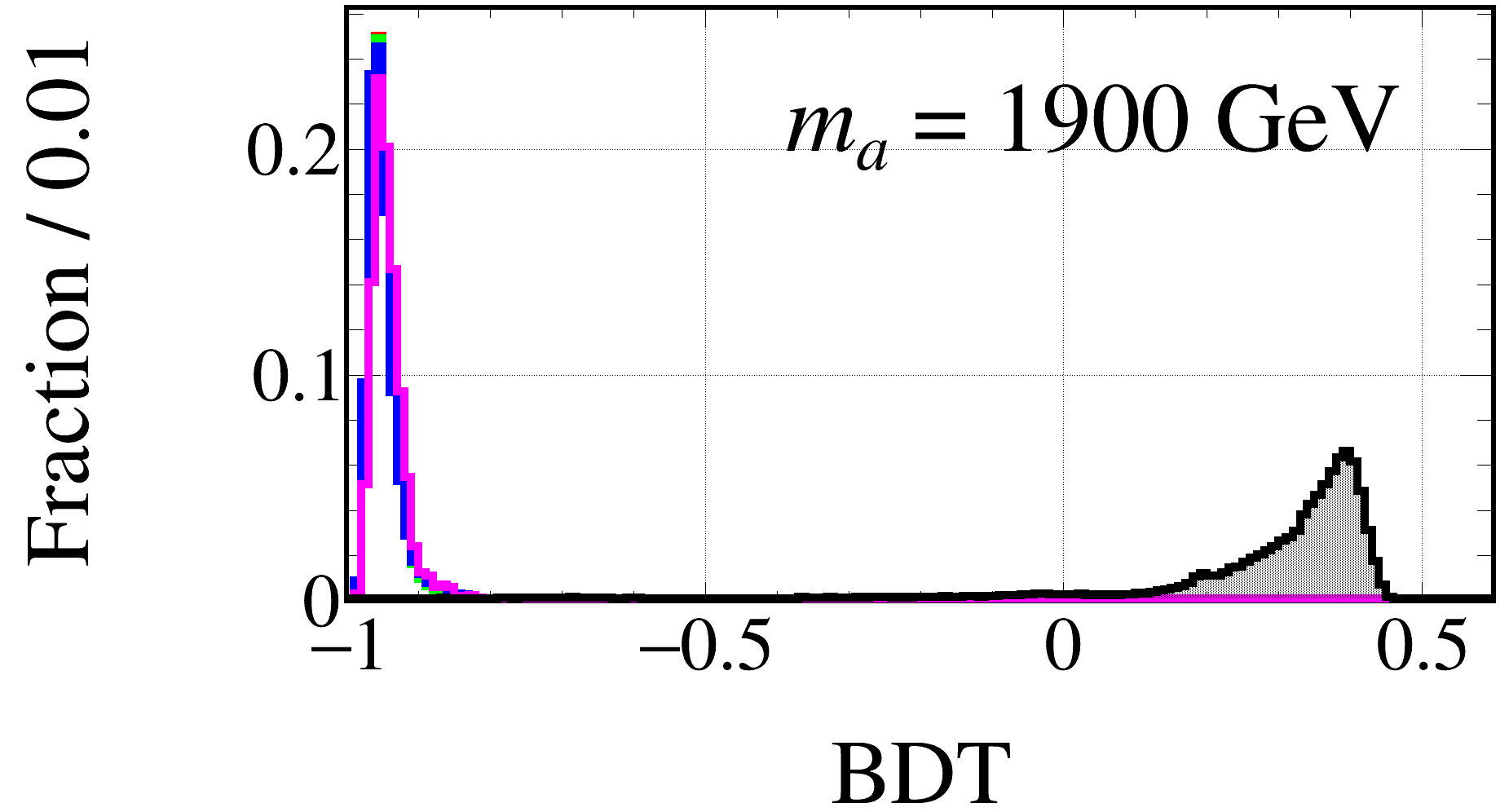}\,\,\,\,\,\,
\includegraphics[width=4.5cm,height=3.5cm]{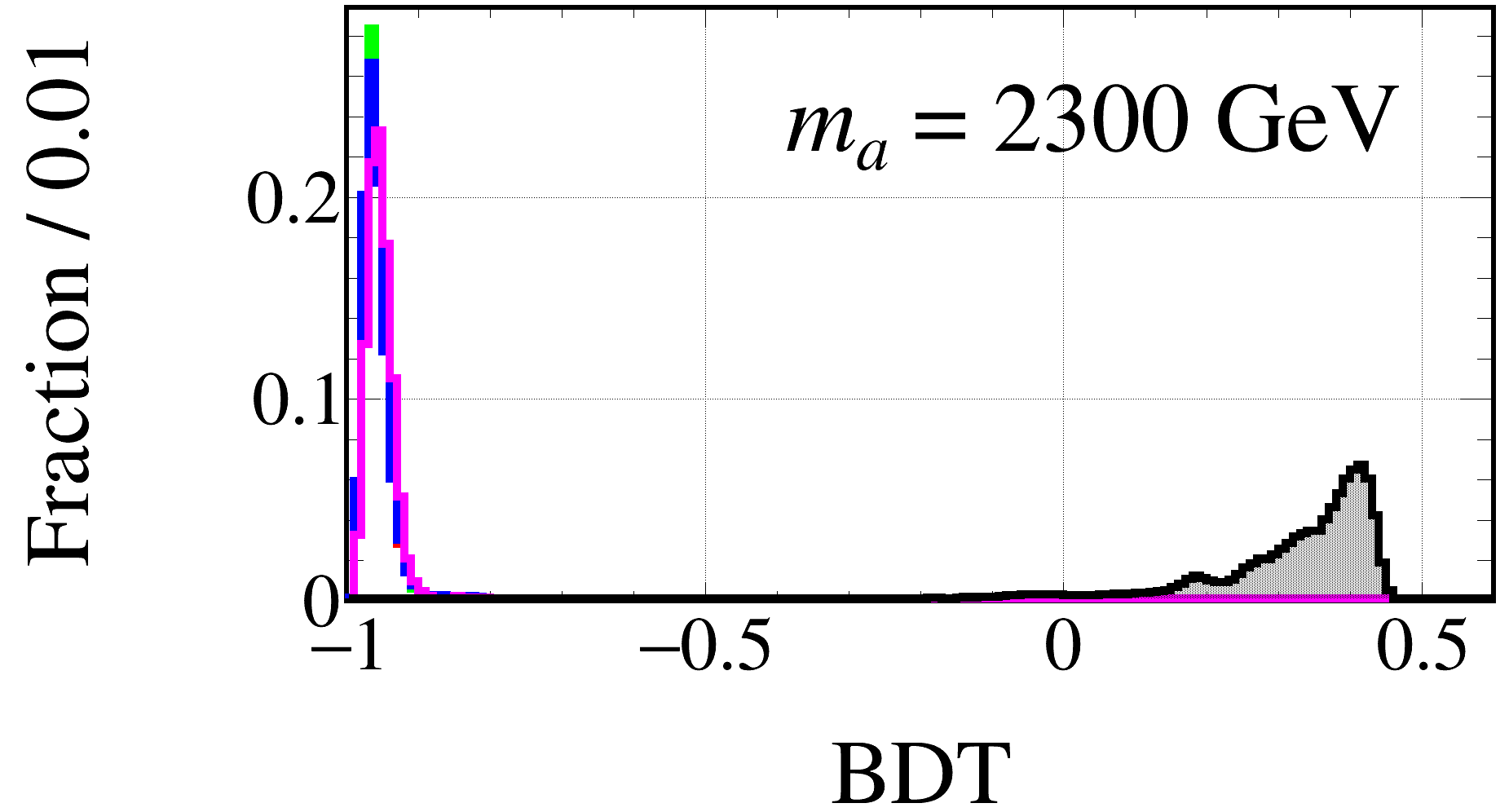}
\caption{
BDT response distributions for the signal (black, filled) with fixed parameter $C_{\tilde{G}}/f_a = 10^{-2} \,\,  {\rm TeV^{-1}}$ and dominant SM background processes at the HL-LHC assuming different $m_a$ cases.
}
\label{fig:BDT_Dis}
\end{figure}

\newpage
\section{Distributions of observables for different ALP masses}
\label{append:Obsrank}

\begin{figure}[h]
\centering
\includegraphics[width=4.5cm,height=3.5cm]{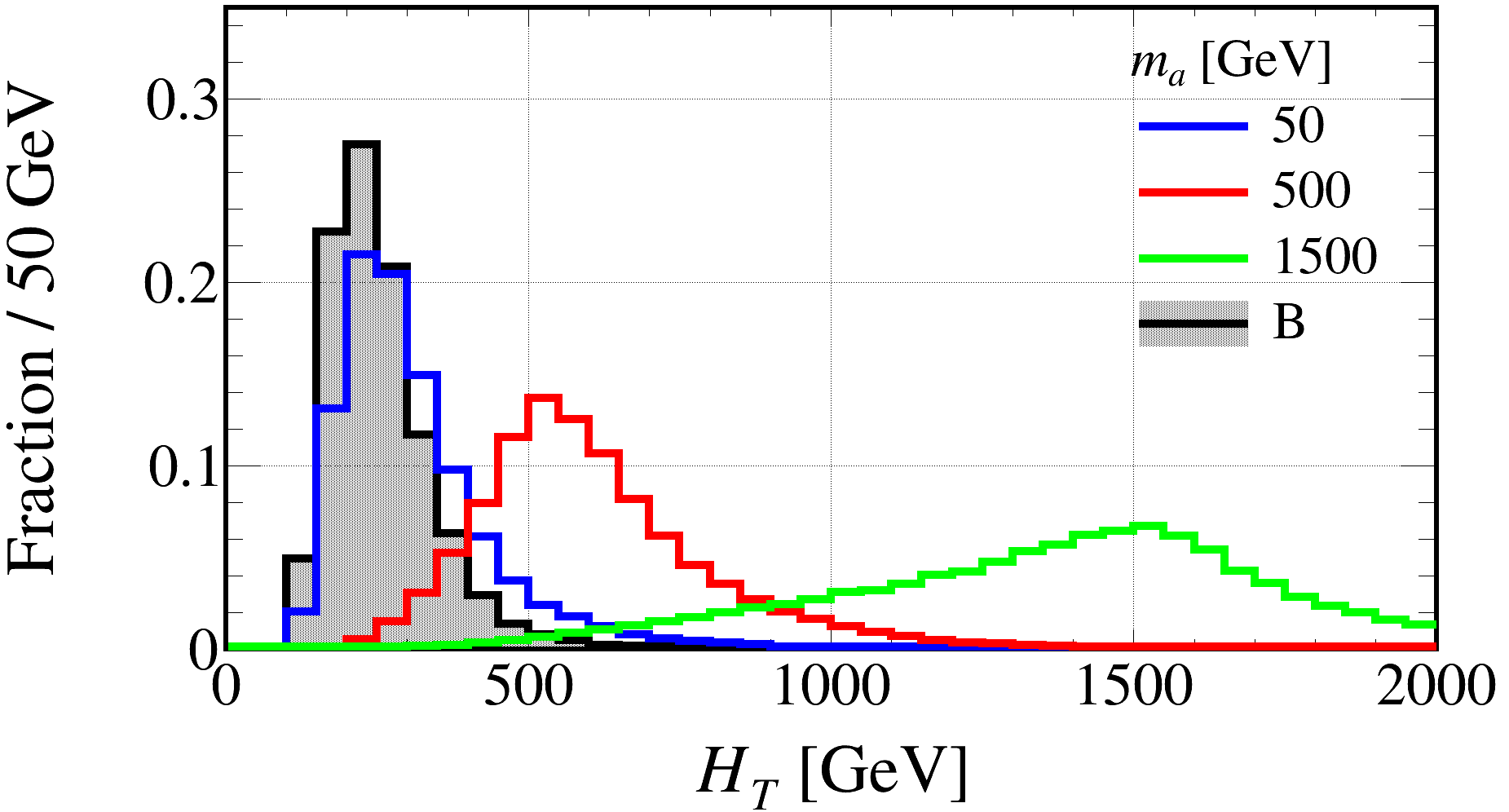}\,\,\,\,\,\,
\includegraphics[width=4.5cm,height=3.5cm]{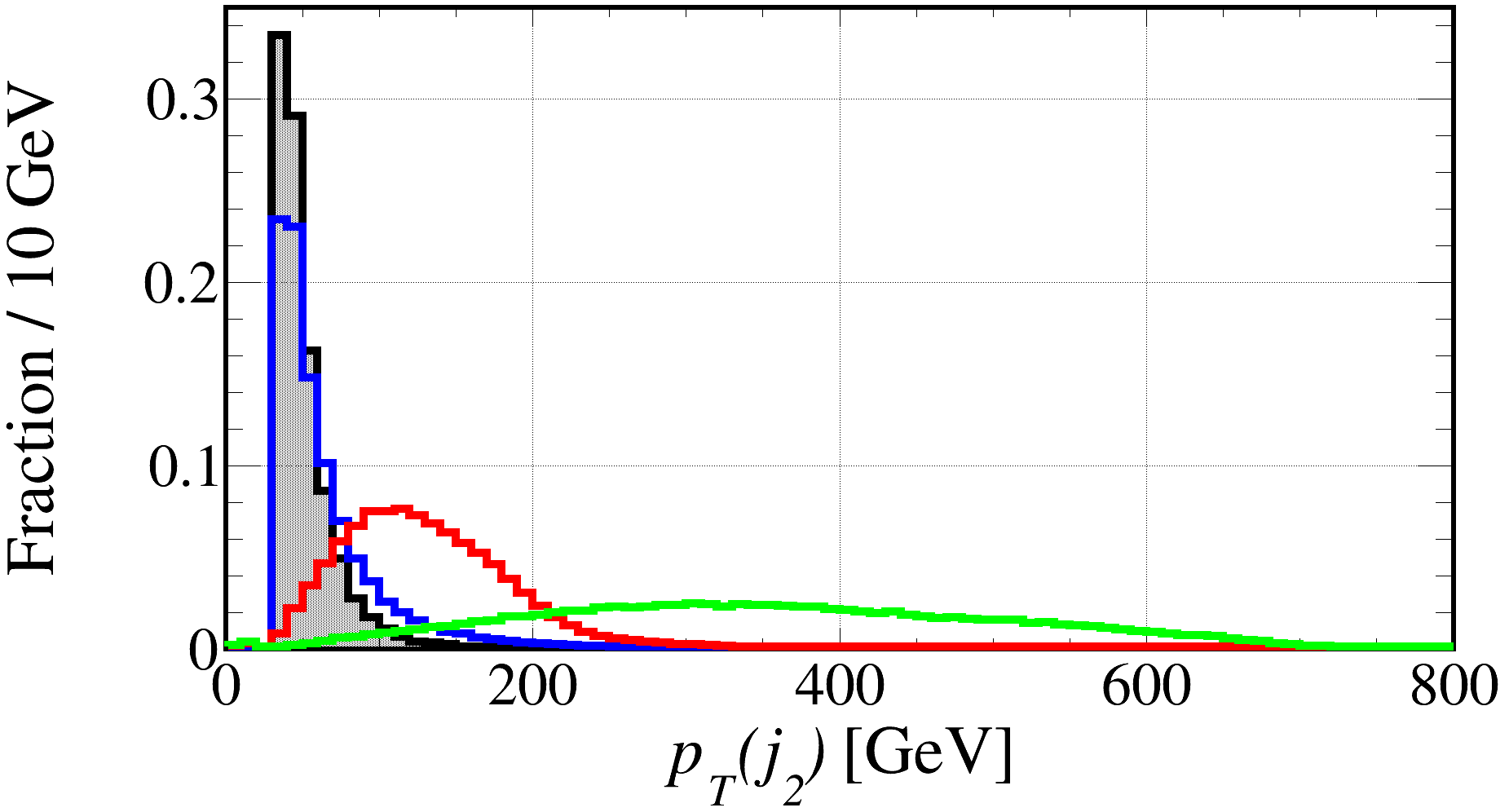}\,\,\,\,\,\,
\includegraphics[width=4.5cm,height=3.5cm]{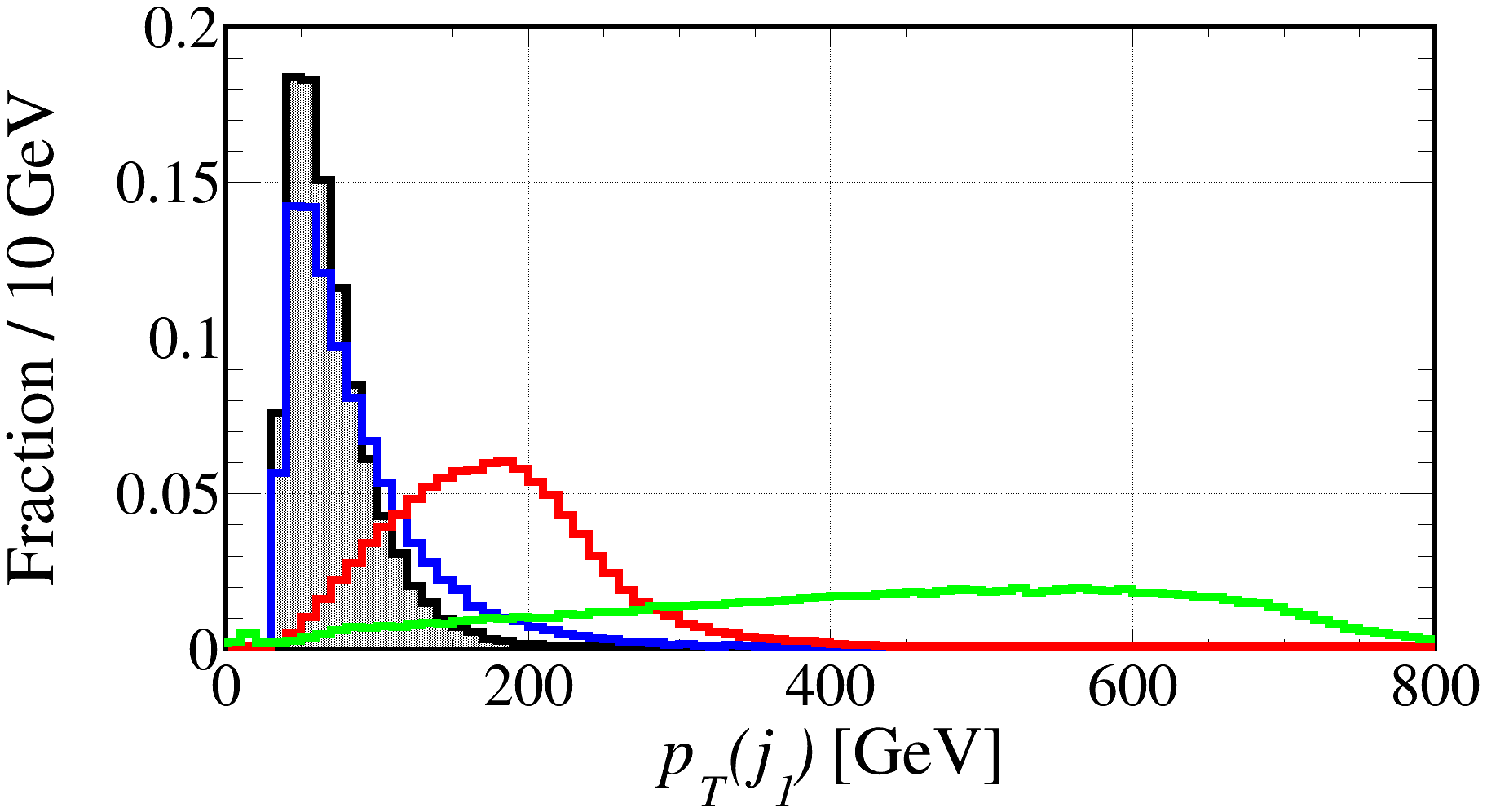}
\par\medskip
\includegraphics[width=4.5cm,height=3.5cm]{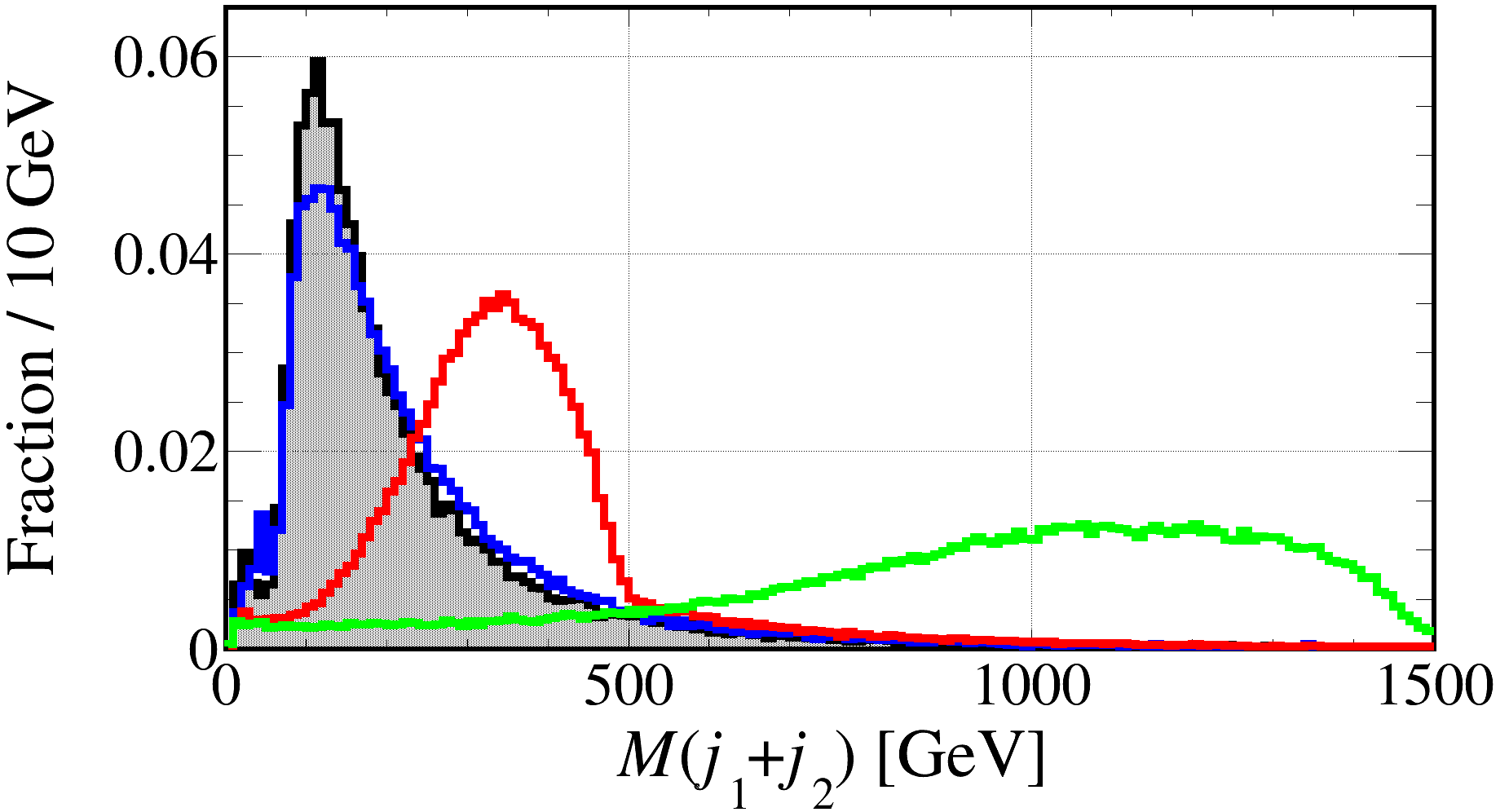}\,\,\,\,\,\,
\includegraphics[width=4.5cm,height=3.5cm]{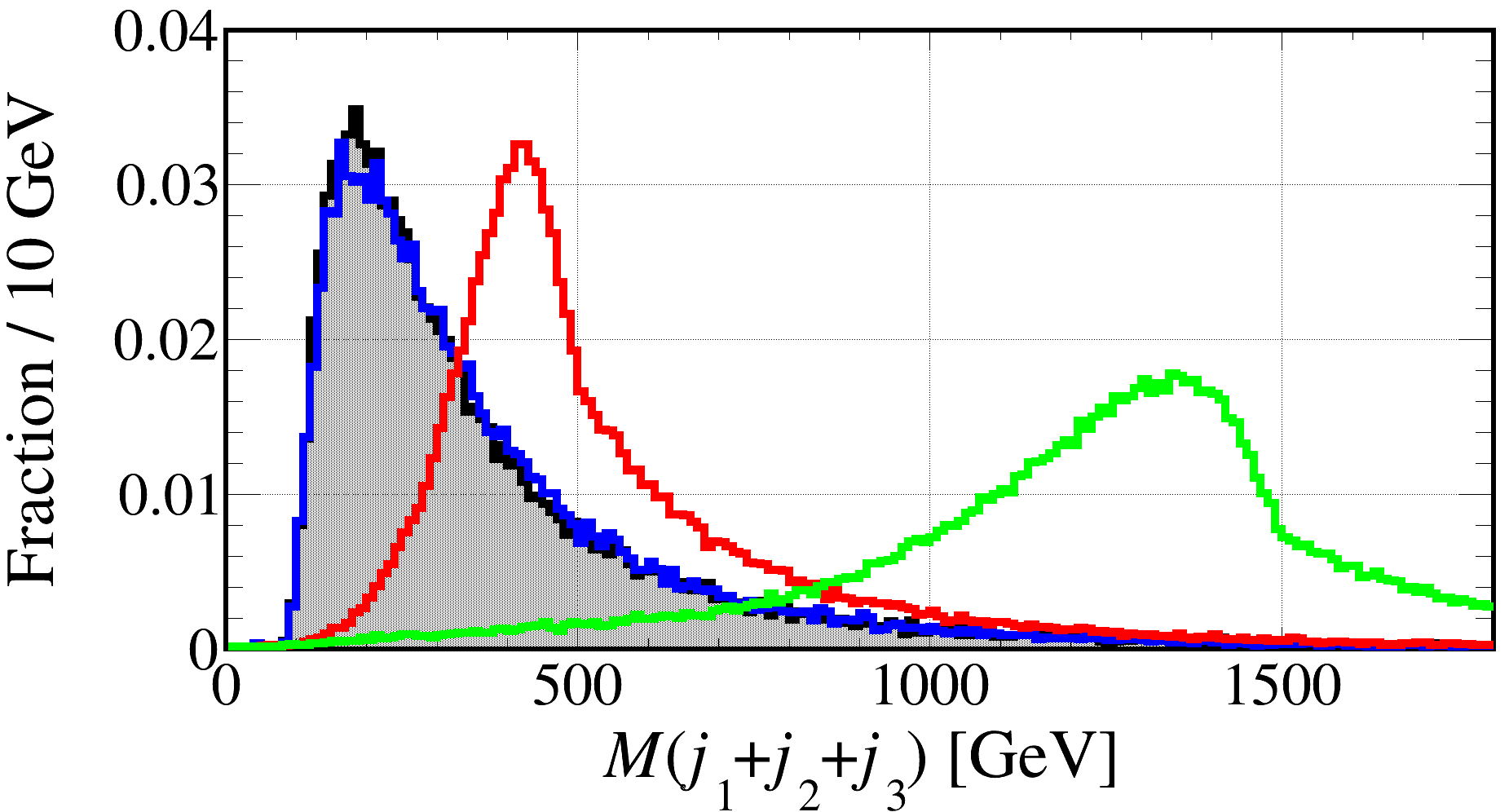}\,\,\,\,\,\,
\includegraphics[width=4.5cm,height=3.5cm]{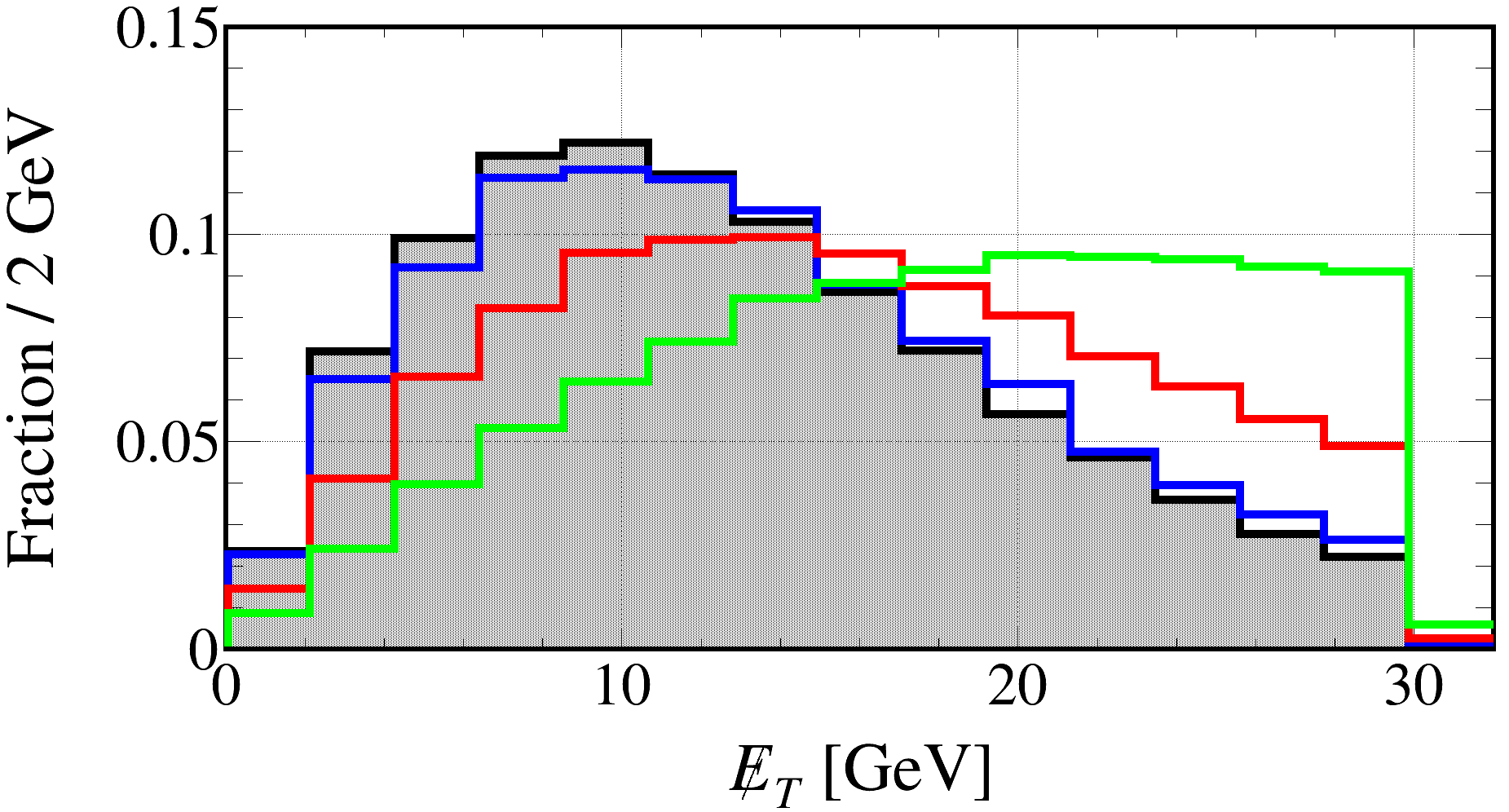}
\par\medskip
\caption{
Distributions of observables for the signal with benchmark coupling $C_{\tilde{G}}/f_a = 10^{-2} \,\, {\rm TeV^{-1}}$ and representative ALP masses, and for the total SM background.
}
\label{fig:Input_Disc_Var_masses}
\end{figure}

\newpage
\section{Selection Efficiency Table}
\label{append:Sel_Eff}

\begin{table}[ht]
\centering
\resizebox{\textwidth}{!}{
\begin{tabular}{ccccccccccc}
\hline
\hline
$m_a$ & cut & signal & multi-jet & $W^\pm j$ & $Z j$ & $\tau^+ \tau^- j$ & $t\bar{t}$ & $W^+ W^-$ & $ZZ$ \\
\hline
\multirow{2}{*}{5 GeV}    & pre-selection & $8.04\mltp10^{-3}$ & $3.14\mltp10^{-4}$ & $6.93\mltp10^{-2}$ & $7.79\mltp10^{-2}$ & $2.18\mltp10^{-3}$ & $5.94\mltp10^{-2}$ & $3.01\mltp10^{-1}$ & $2.59\mltp10^{-1}$ \\
& BDT$>$-0.930  & $9.83\mltp10^{-1}$ & $9.57\mltp10^{-1}$ & $9.59\mltp10^{-1}$ & $9.60\mltp10^{-1}$ & $9.81\mltp10^{-1}$ & $9.84\mltp10^{-1}$ & $9.65\mltp10^{-1}$ & $9.68\mltp10^{-1}$ \\
\hline
\multirow{2}{*}{50 GeV}   & pre-selection & $3.69\mltp10^{-2}$ & $3.14\mltp10^{-4}$ & $6.93\mltp10^{-2}$ & $7.79\mltp10^{-2}$ & $2.18\mltp10^{-3}$ & $5.94\mltp10^{-2}$ & $3.01\mltp10^{-1}$ & $2.59\mltp10^{-1}$ \\
& BDT$>$-0.854  & $3.63\mltp10^{-1}$ & $2.22\mltp10^{-1}$ & $2.55\mltp10^{-1}$ & $2.55\mltp10^{-1}$ & $3.14\mltp10^{-1}$ & $7.16\mltp10^{-1}$ & $3.64\mltp10^{-1}$ & $3.85\mltp10^{-1}$ \\
\hline
\multirow{2}{*}{100 GeV}  & pre-selection & $1.35\mltp10^{-1}$ & $3.14\mltp10^{-4}$ & $6.93\mltp10^{-2}$ & $7.79\mltp10^{-2}$ & $2.18\mltp10^{-3}$ & $5.94\mltp10^{-2}$ & $3.01\mltp10^{-1}$ & $2.59\mltp10^{-1}$ \\
& BDT$>$-0.865  & $2.08\mltp10^{-1}$ & $4.08\mltp10^{-1}$ & $5.06\mltp10^{-1}$ & $5.06\mltp10^{-1}$ & $5.43\mltp10^{-1}$ & $8.53\mltp10^{-1}$ & $5.81\mltp10^{-1}$ & $6.17\mltp10^{-1}$ \\
\hline
\multirow{2}{*}{200 GeV}  & pre-selection & $3.66\mltp10^{-1}$ & $3.14\mltp10^{-4}$ & $6.93\mltp10^{-2}$ & $7.79\mltp10^{-2}$ & $2.18\mltp10^{-3}$ & $5.94\mltp10^{-2}$ & $3.01\mltp10^{-1}$ & $2.59\mltp10^{-1}$ \\
& BDT$>$-0.836  & $1.41\mltp10^{-1}$ & $1.39\mltp10^{-1}$ & $1.59\mltp10^{-1}$ & $1.61\mltp10^{-1}$ & $1.39\mltp10^{-1}$ & $7.80\mltp10^{-1}$ & $3.02\mltp10^{-1}$ & $3.37\mltp10^{-1}$ \\
\hline
\multirow{2}{*}{300 GeV}  & pre-selection & $4.54\mltp10^{-1}$ & $3.14\mltp10^{-4}$ & $6.93\mltp10^{-2}$ & $7.79\mltp10^{-2}$ & $2.18\mltp10^{-3}$ & $5.94\mltp10^{-2}$ & $3.01\mltp10^{-1}$ & $2.59\mltp10^{-1}$ \\
& BDT$>$-0.759  & $1.08\mltp10^{-1}$ & $3.99\mltp10^{-2}$ & $4.45\mltp10^{-2}$ & $4.10\mltp10^{-2}$ & $2.48\mltp10^{-2}$ & $4.57\mltp10^{-1}$ & $1.28\mltp10^{-1}$ & $1.32\mltp10^{-1}$ \\
\hline
\multirow{2}{*}{500 GeV}  & pre-selection & $4.38\mltp10^{-1}$ & $3.14\mltp10^{-4}$ & $6.93\mltp10^{-2}$ & $7.79\mltp10^{-2}$ & $2.18\mltp10^{-3}$ & $5.94\mltp10^{-2}$ & $3.01\mltp10^{-1}$ & $2.59\mltp10^{-1}$ \\
& BDT$>$-0.865  & $2.07\mltp10^{-1}$ & $7.33\mltp10^{-2}$ & $6.55\mltp10^{-2}$ & $6.27\mltp10^{-2}$ & $4.91\mltp10^{-2}$ & $5.74\mltp10^{-1}$ & $1.62\mltp10^{-1}$ & $1.68\mltp10^{-1}$ \\
\hline
\multirow{2}{*}{700 GeV}  & pre-selection & $3.76\mltp10^{-1}$ & $3.14\mltp10^{-4}$ & $6.93\mltp10^{-2}$ & $7.79\mltp10^{-2}$ & $2.18\mltp10^{-3}$ & $5.94\mltp10^{-2}$ & $3.01\mltp10^{-1}$ & $2.59\mltp10^{-1}$ \\
& BDT$>$-0.199  & $1.13\mltp10^{-1}$ & $7.28\mltp10^{-4}$ & $1.44\mltp10^{-3}$ & $1.16\mltp10^{-3}$ & $9.18\mltp10^{-4}$ & $1.47\mltp10^{-2}$ & $7.15\mltp10^{-3}$ & $8.15\mltp10^{-3}$ \\
\hline
\multirow{2}{*}{900 GeV}  & pre-selection & $3.17\mltp10^{-1}$ & $3.14\mltp10^{-4}$ & $6.93\mltp10^{-2}$ & $7.79\mltp10^{-2}$ & $2.18\mltp10^{-3}$ & $5.94\mltp10^{-2}$ & $3.01\mltp10^{-1}$ & $2.59\mltp10^{-1}$ \\
& BDT$>$-0.209  & $2.20\mltp10^{-1}$ & $7.28\mltp10^{-4}$ & $1.30\mltp10^{-3}$ & $8.99\mltp10^{-4}$ & $9.18\mltp10^{-4}$ & $1.48\mltp10^{-2}$ & $6.32\mltp10^{-3}$ & $7.14\mltp10^{-3}$ \\
\hline
\multirow{2}{*}{1100 GeV} & pre-selection & $2.68\mltp10^{-1}$ & $3.14\mltp10^{-4}$ & $6.93\mltp10^{-2}$ & $7.79\mltp10^{-2}$ & $2.18\mltp10^{-3}$ & $5.94\mltp10^{-2}$ & $3.01\mltp10^{-1}$ & $2.59\mltp10^{-1}$ \\
& BDT$>$-0.222  & $3.05\mltp10^{-1}$ & $7.28\mltp10^{-4}$ & $1.11\mltp10^{-3}$ & $8.13\mltp10^{-4}$ & $9.18\mltp10^{-4}$ & $1.52\mltp10^{-2}$ & $5.82\mltp10^{-3}$ & $6.60\mltp10^{-3}$ \\
\hline
\multirow{2}{*}{1500 GeV} & pre-selection & $1.94\mltp10^{-1}$ & $3.14\mltp10^{-4}$ & $6.93\mltp10^{-2}$ & $7.79\mltp10^{-2}$ & $2.18\mltp10^{-3}$ & $5.94\mltp10^{-2}$ & $3.01\mltp10^{-1}$ & $2.59\mltp10^{-1}$ \\
& BDT$>$-0.358  & $4.70\mltp10^{-1}$ & $7.28\mltp10^{-4}$ & $9.14\mltp10^{-4}$ & $7.28\mltp10^{-4}$ & $9.18\mltp10^{-4}$ & $1.25\mltp10^{-2}$ & $5.52\mltp10^{-3}$ & $6.25\mltp10^{-3}$ \\
\hline
\multirow{2}{*}{1900 GeV} & pre-selection & $1.44\mltp10^{-1}$ & $3.14\mltp10^{-4}$ & $6.93\mltp10^{-2}$ & $7.79\mltp10^{-2}$ & $2.18\mltp10^{-3}$ & $5.94\mltp10^{-2}$ & $3.01\mltp10^{-1}$ & $2.59\mltp10^{-1}$ \\
& BDT$>$-0.457  & $6.60\mltp10^{-1}$ & $7.28\mltp10^{-4}$ & $1.11\mltp10^{-3}$ & $8.56\mltp10^{-4}$ & $4.59\mltp10^{-4}$ & $1.57\mltp10^{-2}$ & $5.85\mltp10^{-3}$ & $6.29\mltp10^{-3}$ \\
\hline
\multirow{2}{*}{2300 GeV} & pre-selection & $1.12\mltp10^{-1}$ & $3.14\mltp10^{-4}$ & $6.93\mltp10^{-2}$ & $7.79\mltp10^{-2}$ & $2.18\mltp10^{-3}$ & $5.94\mltp10^{-2}$ & $3.01\mltp10^{-1}$ & $2.59\mltp10^{-1}$ \\
& BDT$>$-0.582  & $8.71\mltp10^{-1}$ & $7.52\mltp10^{-4}$ & $1.40\mltp10^{-3}$ & $9.42\mltp10^{-4}$ & $9.18\mltp10^{-4}$ & $1.67\mltp10^{-2}$ & $5.95\mltp10^{-3}$ & $6.76\mltp10^{-3}$ \\ 
\hline
\hline                                       
\end{tabular}}
\caption{
Selection efficiencies for the signal processes with a range of representative ALP masses and fixed $C_{\tilde{G}}/f_a = 10^{-2} \,\, {\rm TeV^{-1}}$ and the background processes 
multi-jet, $W^\pm j$, $Z j$, $\tau^+\tau^-j$, $t\bar{t}$, $W^+ W^-$ and $Z Z$  at the HL-LHC with center-of-mass energy $\sqrt{s}=14$ TeV.
}
\label{tab:All_Eff}
\end{table}




\bibliography{Refs}
\bibliographystyle{hep_v5}

\end{document}